\documentclass[12pt]{article}
\usepackage{amsmath,epsfig,graphicx,verbatim}
\usepackage{epstopdf}
\usepackage{color}
\usepackage{subfigure}

\include{epsf}
\def\ltap{\raisebox{-.6ex}{\rlap{$\,\sim\,$}} \raisebox{.4ex}{$\,<\,$}}
\def\gtap{\raisebox{-.6ex}{\rlap{$\,\sim\,$}} \raisebox{.4ex}{$\,>\,$}}

\newcommand\as{\alpha_{\mathrm{S}}}
\newcommand\f[2]{\frac{#1}{#2}}
\def\beq{\begin{equation}}
\def\eeq{\end{equation}}
\def\beeq{\begin{eqnarray}}
\def\eeeq{\end{eqnarray}}

\def\to{\rightarrow}

\def\nn{\nonumber}

\def\tL{{\tilde L}}
\def\msbar{{\overline {\rm MS}}}

\begin{document}
\begin{titlepage}
\renewcommand{\thefootnote}{\fnsymbol{footnote}}
\begin{flushright}
     ZU-TH 20/18
     \end{flushright}
\par \vspace{10mm}

\begin{center}
{\Large \bf Transverse-momentum resummation for\\[0.3cm] top-quark pair 
production at the LHC}
\end{center}

\par \vspace{2mm}
\begin{center}
{\bf Stefano Catani${}^{(a)},$ Massimiliano Grazzini${}^{(b)}$}
and {\bf Hayk Sargsyan$^{(b)}$\\
}

\vspace{5mm}

$^{(a)}$ INFN, Sezione di Firenze and Dipartimento di Fisica e Astronomia,\\
Universit\`a di Firenze,
I-50019 Sesto Fiorentino, Florence, Italy

$^{(b)}$ Physik-Institut, Universit\"at Z\"urich,
CH-8057 Zurich, Switzerland

\vspace{5mm}

\end{center}

\par \vspace{2mm}
\begin{center} {\large \bf Abstract} \end{center}
\begin{quote}
\pretolerance 10000

We consider transverse-momentum resummation for top-quark pair production
in hadron collisions. At small transverse momenta of the top-quark pair,
the logarithmically-enhanced QCD contributions are resummed to all orders up
to next-to-leading logarithmic accuracy. At intermediate and large values of
transverse momenta, the resummation is consistently combined with the complete
result at fixed perturbative order.
We present numerical results for the transverse-momentum distribution of
top-quark pairs at LHC energies.
We perform a detailed study of the scale dependence of the results to
estimate their perturbative uncertainty.
We comment on the comparison with ATLAS and CMS data.

\end{quote}

\vspace*{\fill}
\begin{flushleft}
June 2018

\end{flushleft}
\end{titlepage}

\setcounter{footnote}{1}
\renewcommand{\thefootnote}{\fnsymbol{footnote}}
\section{Introduction}
\label{sec:intro}

The top quark ($t$)
\cite{Olive:2016xmw} is the heaviest known elementary particle. Since
its mass is close to the scale of electroweak symmetry breaking,
the top quark 
is crucial to the hierarchy problem.
Within the Standard Model (SM)
the main source of top-quark events in high-energy hadron collisions is 
the production of top-quark pairs.
The study of top-quark pair ($t{\bar t}$) production at hadron colliders
can shed light on the nature of the electroweak-symmetry breaking.
Moreover, such a study also gives relevant information on the SM backgrounds of many new-physics models.

The theoretical efforts for obtaining accurate predictions for 
$t{\bar t}$
production at hadron colliders started three decades ago with
the calculation of the next-to-leading order (NLO) QCD corrections
to the total cross section
\cite{Nason:1987xz, Beenakker:1988bq, Nason:1989zy}
and kinematical distributions \cite{Mangano:1991jk}
for this production process.
The calculation of the next-to-next-to-leading order (NNLO) QCD
corrections to the $t{\bar t}$ total cross section was completed
\cite{Baernreuther:2012ws} in recent years.
Besides the total cross section,  NNLO predictions for differential cross sections and more general
kinematical distributions are now available \cite{Czakon:2014xsa,Czakon:2016ckf,Czakon:2017dip} and are of
great importance for precision studies.

This paper is devoted to the transverse-momentum 
spectrum
of the top-quark pair.
This observable has been measured by the ATLAS and CMS experiments at the LHC
in proton--proton collisions at the centre--of--mass energies 
$\sqrt{s}=7$~TeV \cite{Chatrchyan:2012saa, Aad:2015eia, Aaboud:2016iot} and
8~TeV \cite{Aaboud:2016iot, Khachatryan:2015oqa, Aad:2015mbv, Sirunyan:2017azo}.
First results at $\sqrt{s}=13$~TeV \cite{Khachatryan:2016mnb, Aaboud:2016syx,
Sirunyan:2017mzl, Sirunyan:2018wem} 
started to appear recently.

The bulk of the cross section is produced in the
kinematical region where the transverse momentum $q_T$ of the top-quark pair is
smaller
than the mass $m_t$ of the top quark.
The $q_T$ differential cross section $d\sigma/dq_T$ of the top-quark pair is 
computable in QCD
perturbation theory
\cite{Mangano:1991jk, Dittmaier:2007wz, Melnikov:2010iu}. In the small-$q_T$ region $(q_T \ll m_t)$, 
the perturbative expansion
is badly behaved,
since the size of the perturbative coefficients is enhanced by powers of
$\ln q_T$.
A reliable theoretical calculation of the detailed $q_T$ shape of 
$d\sigma/dq_T$
requires the all-order resummation of these
logarithmically-enhanced terms. 

This type of perturbative behaviour at small $q_T$ is well
known
\cite{Dokshitzer:hw, Parisi:1979se, Curci:1979bg}
from the 
hadroproduction process of a high-mass lepton pair through
the Drell--Yan (DY) mechanism. In the case of the DY process the
all-order resummation
of the $\ln q_T$ terms is fully understood
\cite{Dokshitzer:hw, Parisi:1979se, Curci:1979bg, Collins:1984kg}.
At the level of leading-logarithmic (LL) contributions, the extension of
resummation from the DY process to the heavy-quark process 
is relatively straightforward \cite{Berger:1993yp,Mrenna:1996cz}.
However, beyond the LL level the structure of $\ln q_T$ terms for the
heavy-quark process
is definitely different
from that of the DY process.
Indeed, transverse-momentum
resummation for the DY process, and, more generally, for the production
of {\em colourless} high-mass systems,
has an all-order {\em universal} 
(process-independent) structure.
This universality structure eventually originates from the fact that
the transverse momentum of the colourless system is produced by (soft and
collinear) QCD
radiation from the initial-state colliding partons. The heavy-quark
production process
definitely belongs to a different class of processes, since the produced
final-state heavy quarks carry colour charge and, therefore, they act as
additional
source of QCD radiation. The transverse momentum of the heavy-quark pair
depends on initial-state
radiation, on final-state radiation and on quantum (and colour flow)
interferences between radiation from the initial and final states.
These physical differences
lead to conceptual and technical complications in the theoretical formulation
of transverse-momentum resummation for heavy-quark production.

The all-order $q_T$ resummation for the heavy-quark production process 
was discussed in Refs.~\cite{Zhu:2012ts, Li:2013mia, Catani:2014qha}.
The analysis of Refs.~\cite{Zhu:2012ts, Li:2013mia} is based on 
Soft Collinear Effective Theory (SCET),
and it is limited to the treatment of the $q_T$ cross section averaged over 
the azimuthal angles of the produced heavy quarks.
Corresponding quantitative results for $t{\bar t}$ production at the LHC were obtained in Refs.~\cite{Zhu:2012ts, Li:2013mia} by following and extending the implementation formalism of Ref.~\cite{Becher:2011xn}.
An independent formulation of transverse-momentum resummation for 
heavy-quark pairs, including
the complete treatment of azimuthal correlations is presented
in Ref.~\cite{Catani:2014qha}. A main difference between the production of 
heavy quarks and of colourless systems is the appearance of a new resummation
factor that is due to soft-parton radiation at 
wide
angles with respect to 
the direction of the colliding hadrons. This additional resummation factor
embodies the effect of soft radiation from the heavy-quark final state and 
from initial-state and final-state interferences.
This factor is controlled by a soft anomalous dimension that depends on 
the angular (rapidity) distribution of the produced heavy quark and antiquark
and on the colour configuration of their underlying production mechanism.

In this paper we present
quantitative results for the
transverse-momentum spectrum of a top-quark pair 
at the LHC. The results are obtained by using the all-order
resummation formalism of Ref.~\cite{Catani:2014qha}, which is directly implemented in impact parameter space \cite{Bozzi:2005wk}. 
We perform the resummation 
up to the next-to-leading logarithmic (NLL) level, by also explicitly
including all (logarithmic and non logarithmic) contributions up to NLO
in the perturbative QCD expansion. We study the perturbative uncertainty 
of our results
and compare them to the available LHC data at $\sqrt{s}=8$~TeV
\cite{Aaboud:2016iot, Khachatryan:2015oqa}.
We also study the quantitative impact of the soft wide-angle factor 
that is a distinctive feature of
heavy-quark production. Finally we present a comparison of our resummed results 
for the $t{\bar t}$ spectrum with those obtained by using a Monte Carlo generator that interfaces the NLO calculation to the PYTHIA parton shower 
\cite{Sjostrand:2006za}
by using the POWHEG BOX implementation \cite{Alioli:2010xd}.

The quantitative results presented in this paper are limited to the $q_T$ cross section integrated over the azimuthal angles at the $t{\bar t}$ pair.
Corresponding results for azimuthal correlations
of the top-quark pair have been presented in Ref.~\cite{Catani:2017tuc}.
Those results are included in a more general (process-independent) discussion
\cite{Catani:2017tuc} of highly non-trivial features related to azimuthal correlation effects in QCD.

The paper is organized as follows. In Sect.~\ref{sec:theo} we briefly recall the theoretical framework and the resummation formalism that we use to carry out our calculation.
In Sect.~\ref{sec:result} we present our numerical results and comparisons with LHC data. In Sect.~\ref{sec:fo} we consider fixed-order calculations at NLO and NNLO,
and then in Sect.~\ref{sec:res} we present our resummed predictions.
Our results are summarized in Sect.~\ref{sec:sum}.

\section{Transverse-momentum resummation}
\label{sec:theo}

The resummation formalism that we use in this paper is discussed in detail in 
Ref.~\cite{Catani:2014qha}.
The formalism can be applied to a generic pair of heavy quarks that is produced in hadron--hadron collisions. In this Section we briefly recall the main points of the formalism, by focusing on the specific case of 
the 
transverse-momentum spectrum of a $t{\bar t}$ pair.

We consider the inclusive hard-scattering process $h_1+h_2 \to t{\bar t} +X$, 
where the two colliding hadrons $h_1$ and $h_2$ with centre--of--mass energy
$\sqrt s$ produce the $t{\bar t}$ pair, and $X$ denotes the accompanying final-state radiation. Although we are mostly interested in the single-differential cross section
$d\sigma/dq_T$, in our presentation we consider the triple-differential cross section with respect to $q_T$, the invariant mass $M$ of the $t{\bar t}$ pair and the scattering angle of the quark or antiquark. Indeed, the use of the angular dependent cross section clarifies how resummation can be systematically organized in exponential form (see Eqs.~(\ref{eq:ff}) and (\ref{eq:gexp}) and the 
ensuing discussion in this 
Section).
The differential cross section at fixed values of $q_T$, $M$
and of the polar angle $\theta$ of the top quark in the Collins--Soper 
rest frame\footnote{The polar angle in other rest frames of the 
$t{\bar t}$ pair can equivalently be used 
(see Appendix~A in Ref.~\cite{Catani:2015vma}).} 
\cite{Collins:1977iv}
of the $t{\bar t}$ pair can be written as
\begin{align}
\f{d\sigma}{dq_T^2\,dM^2\,d\cos\theta}(q_T,M,s,\theta)&=\sum_{a,b}\int_0^1 dx_1\int_0^1 dx_2 \;f_{a/h_1}(x_1,\mu_F^2) \;f_{b/h_2}(x_2,\mu_F^2)\nn\\
& \times\f{d{\hat \sigma}_{ab}}{dq_T^2\,dM^2\,d\cos\theta}(q_T,M,{\hat s},\theta;\as(\mu_R^2),\mu_R^2,\mu_F^2)\, \;,
\label{eq:triplesigma}
\end{align}
where $a, b$ denotes the parton indices ($a=q,{\bar q},g$),
$f_{a/h}(x,\mu_F^2)$ are the parton distribution functions (PDFs)
of the colliding hadron $h$ at the factorization scale $\mu_F$,
$d{\hat \sigma}_{ab}$ is the partonic differential cross section
for the partonic process $a+b \to t{\bar t} +X$, ${\hat s}=x_1x_2 s$ is the square of the partonic
centre-of-mass energy and $\as(\mu_R^2)$ is the QCD coupling evaluated at the renormalization scale $\mu_R$.
We use the $\msbar$ renormalization scheme for the QCD coupling and the 
$\msbar$ factorization scheme for the PDFs.
The cross sections $d\sigma$ and $d{\hat \sigma}$ obviously depend on the mass
$m_t$ of the top quark, but the $m_t$ dependence is not explicitly denoted
in all the formulae of this Section.



The resummation is performed at the level of the partonic cross section, which is first decomposed as \cite{Catani:2014qha}
\begin{equation}
\label{eq:singreg}
d{\hat \sigma}=d{\hat \sigma}^{\rm (sing)}+d{\hat \sigma}^{\rm (reg)}\, ,
\end{equation}
where, order-by-order in perturbation theory, the component $d{\hat \sigma}^{\rm (sing)}$ embodies {\it all} the singular terms in the $q_T\to 0$ limit,
whereas $d{\hat \sigma}^{\rm (reg)}$ includes the remaining non-singular terms. At the $n$-th order in the expansion in powers of $\as$ we thus have $d{\hat \sigma}^{\rm (reg)}/d{\hat \sigma}^{\rm (sing)}={\cal O}(q_T/M)$ (modulo logarithmic corrections) and $d{\hat \sigma}^{\rm (reg)}$ represents a power suppressed correction (by at least one power of $q_T/M$) to $d{\hat \sigma}^{\rm (sing)}$
as $q_T\to 0$. 

In our resummation treatment the `singular' component $d{\hat \sigma}^{\rm (reg)}$, which contains all the singular logarithmically-enhanced contributions at small $q_T$,
is evaluated by resumming these contributions to all order in $\as$.
The `regular' component $d{\hat \sigma}^{\rm (reg)}$ is instead evaluated at a specified fixed-order accuracy.


The resummation procedure of the logarithmically-enhanced terms in 
$d{\hat \sigma}^{\rm (sing)}$ is carried out in the impact parameter space. 
The impact parameter $\bf b$ is the conjugate variable to ${\bf q}_T$
through a Fourier transformation. The small-$q_T$ region ($q_T \ll M$)
corresponds to the large-$b$ region ($M b \gg 1$) and the logarithmic terms
$\ln(q_T/M)$ become large logarithmic contributions $\ln(M^2b^2)$ in $b$ space.
The resummed component of the partonic cross section is then obtained by performing the inverse Fourier transformation with respect to the impact parameter $b$.
For the azimuthally-integrated \cite{Catani:2014qha, Catani:2017tuc} 
transverse-momentum partonic cross section the Fourier transformation
turns into a Bessel transformation, and the singular component of the partonic cross section in the right-hand side of Eq.~(\ref{eq:triplesigma}) can be written as
\begin{equation}
\label{eq:resw}
\frac{d{\hat \sigma}^{\rm (sing)}_{a_1a_2}}{dq_T^2\,dM^2\,d\cos\theta}
=\as^2 \sum_{c=q,{\bar q},g} 
\frac{d{\hat \sigma}_{c{\bar c}}^{(0)}(M,{\hat s},\theta)}{dM^2 \;d\cos\theta}
\int_0^\infty db \;\f{b}{2} \,J_0(bq_T)\,
{\cal W}_{c{\bar c}\,\leftarrow a_1a_2}(b,M,{\hat s},\theta;\as,\mu_R^2,\mu_F^2)\, ,
\end{equation}
where $J_0(x)$ is the 0th-order Bessel function, and 
$\as^2 \,d{\hat \sigma}_{c{\bar c}}^{(0)}$ is the leading-order (LO) differential cross section for the partonic process $c+{\bar c} \to t {\bar t}$.
As explicitly denoted by the sum over $c$ in Eq.~(\ref{eq:resw}), the LO process only
involves the two partonic channels of quark-antiquark annihilation 
($c{\bar c}=q{\bar q},{\bar q}q$) and gluon fusion ($c{\bar c}=gg$).
The resummation of high-order radiative corrections is embodied in the $b$-space resummed factor
${\cal W}_{c{\bar c}\,\leftarrow a_1a_2}$.

The all-order resummation structure of ${\cal W}_{c{\bar c}\,\leftarrow a_1a_2}$ 
can be organized in exponential form \cite{Catani:2014qha}.
This structure is better expressed by defining \cite{Bozzi:2005wk}
the Mellin $N$-moments ${\cal W}_N(M)$ of ${\cal W}(M,{\hat s})$
with respect to the variable
$z=M^2/{\hat s}$ at fixed $M$. 
For the sake of simplicity, in the 
subsequent
presentation we omit the explicit dependence on the parton indices and also the dependence on colour indices
(we postpone some comments on the colour structure and colour correlations).
The $N$-moment ${\cal W}_{N}$ of the resummed factor ${\cal W}$ in Eq.~(\ref{eq:resw})
has
the form
\begin{equation}
\label{eq:ff}
{\cal W}_{N}(b,M,\theta;\as,\mu_R^2,\mu_F^2)=
{\cal H}_N (M,\theta;\as,\mu_R^2,\mu_F^2,Q)\; 
\exp\{{\cal G}_N(\theta,\as,{\tilde L};M/\mu_R,M/Q)\}\, ,
\end{equation}
where the exponent ${\cal G}$ can be expanded as
\begin{align}
{\cal G}_N(\theta,\as,\tL;M/\mu_R,M/Q)\}&=\tL g^{(1)}(\as \tL)
+g_N^{(2)}(\theta,\as \tL;M/\mu_R,M/Q) \nn \\
&+\sum_{k=3}^\infty \as^{k-2} \;g_N^{(k)}(\theta,\as \tL;M/\mu_R,M/Q) \;\;.
\label{eq:gexp}
\end{align}
The dependence of ${\cal G}$ on the impact parameter $b$ is encoded in the logarithmic expansion parameter
\begin{equation}
\label{eq:tL}
\tL\equiv \ln(Q^2b^2/b_0^2+1) \;\;,
\end{equation}
where $b_0=2e^{-\gamma_E}$ ($\gamma_E=0.5772...$ is the Euler number) and the auxiliary scale $Q$ ($Q\sim M$) is named resummation scale \cite{Bozzi:2003jy}.
Note that the right-hand side of Eq.~(\ref{eq:gexp}) is not a customary order-by-order
expansion in $\as$, but rather an expansion in $\as$ at fixed $\as\tL$. Such expansion consistently resums classes of large logarithmic terms ($\tL \gg 1, \; 
\as\tL \sim {\cal O}(1)$).
The function $g^{(1)}$ controls the LL
contributions $\as^n \tL^{n+1}$ to ${\cal G}_N$,
$g^{(2)}_N$ controls the NLL
contributions $\as^n \tL^{n}$, and so forth.
The resummation scale $Q$ is introduced \cite{Bozzi:2003jy} to parametrize
the arbitrariness in the truncation of the logarithmic expansion in 
Eq.~(\ref{eq:gexp}). Changing the value of $Q$ produces effects on $g^{(k)}_N$ 
($k \geq 2$) that are parametrically of the same order as those due to 
$\as g^{(k+1)}_N$. Therefore, variations of $Q$ around the value $Q=M$ at a given logarithmic accuracy can be used to estimate uncertainties due to neglected subdominant classes of logarithmic contributions.
The function ${\cal H}$ in Eq.~(\ref{eq:ff})
does not depend on the impact parameter $b$ and, therefore, it can
be expanded in powers of $\as(\mu_R^2)$:
\begin{equation}
%
{\cal H}_N
={\cal H}_N^{(0)}
+\left(\f{\as}{\pi}\right)\,{\cal H}_N^{(1)}+
\left(\f{\as}{\pi}\right)^2\,{\cal H}_N^{(2)}+{\cal O}(\as^3)\, ,
\label{eq:hfun}
\end{equation}
where each perturbative term ${\cal H}_N^{(k)}$ depends on $M,\theta$ and the auxiliary scales $\mu_R,\mu_F,Q$.



Since the factor ${\cal H}_N$ in Eq.~(\ref{eq:ff}) does not contain logarithmically-enhanced contributions, we treat it by truncation of the perturbative expansion in Eq.~(\ref{eq:hfun})
at a given fixed-order accuracy, analogously to the treatment of the regular component $d{\hat \sigma}^{\rm (reg)}$ of the cross section in Eq.~(\ref{eq:singreg}).


The computation of $d{\hat \sigma}^{\rm (sing)}$ and $d{\hat \sigma}^{\rm (reg)}$ at a given logarithmic and fixed-order accuracy, respectively,
defines a systematic expansion \cite{Bozzi:2005wk} of Eq.~(\ref{eq:singreg}).
This expansion can be used to obtain predictions that contain the full information of the perturbative calculation up to a given fixed order plus resummation of the logarithmically-enhanced contributions at small $q_T$. The various orders of this expansion are denoted by LL, NLL+NLO, NNLL+NNLO and so forth, where the first label (LL, NLL, NNLL, ...) denotes the logarithmic accuracy, while the second label (NLO, NNLO, ...) refers to the corresponding perturbative order of the 
total cross section. In particular, the NLL+NLO expansion of Eq.~(\ref{eq:singreg})
is obtained by including the functions $g^{(1)}$ and $g^{(2)}_N$ and the coefficients ${\cal H}^{(1)}$ in the resummed component, and by computing the finite component 
$d{\hat \sigma}^{\rm (reg)}$
at ${\cal O}(\as^3)$.

We 
remark
that the NLL+NLO (NNLL+NNLO) result includes the full NLO (NNLO) perturbative contribution in the small-$q_T$ region.
Considering a generic upper limit value $q_{T \rm max}$, the integral over 
the range $0 \leq q_T \leq q_{T \rm max}$ of the $q_T$ differential cross section
at the NLL+NLO (NNLL+NNLO) level
includes the complete perturbative terms up to NLO (NNLO).
In particular, 
since the logarithmic variable $\tL$ in Eq.~(\ref{eq:tL}) vanishes at $b=0$,
the NLO (NNLO)
total ($q_T$ integrated) cross section $d\sigma/dM^2\,d\cos\theta$ is {\em exactly} recovered \cite{Bozzi:2003jy, Bozzi:2005wk} upon integration over $q_T$
of the NLL+NLO (NNLL+NNLO) result.
In the case of heavy-quark production, the coefficient ${\cal H}^{(2)}$ is
still unknown: this prevents us from performing calculations at NNLL+NNLO accuracy. In the following sections of this paper we thus limit ourselves to presenting 
resummed results at NLL+NLO accuracy.

In Eqs.~(\ref{eq:triplesigma})--(\ref{eq:hfun}) and accompanying comments we have briefly summarized the main results of Ref.~\cite{Catani:2014qha} on $q_T$ resummation for $t{\bar t}$ production. In this presentation we have used a notation and a style that closely follow analogous results \cite{Bozzi:2005wk} for production processes of colourless high-mass systems
(e.g., the DY process or Higgs boson production). As discussed in 
Ref.~\cite{Catani:2014qha} and recalled in Sect.~\ref{sec:intro},
there are important differences between $t{\bar t}$ production and the 
production of colourless systems. In the following we highlight more explicitly these differences.

The LL function $g^{(1)}$ in Eq.~(\ref{eq:gexp}) is completely analogous to the corresponding function for colourless production. More precisely, taking into account the dependence on the parton indices (which has been neglected in the notation of Eqs.~(\ref{eq:ff}) and (\ref{eq:gexp})), $g^{(1)}$ depends on the flavour of the partons (see Eq.~(\ref{eq:resw})) but it is flavour diagonal. The function 
$g^{(1)}$ indeed depends on the partonic channel that produces the $t{\bar t}$ pair at
the LO level. Since $t{\bar t}$ production receives LO contributions from $q{\bar q}$ annihilation and $gg$ fusion, both channels have to be considered. The 
function $g^{(1)}$ for $t{\bar t}$ production exactly coincides with the corresponding function for DY and Higgs boson production in the $q{\bar q}$
and $gg$ channel, respectively.

Analogously to the production of colourless systems, the functions
$g^{(k)}_N$ (with $k\geq 2$) in Eq.~(\ref{eq:gexp}) are matrices in the flavour indices of the partons. Therefore, the practical implementation of the resummation formula (\ref{eq:gexp}) in exponentiated form requires a proper diagonalization of these matrices with respect to the flavour indices. Such diagonalization and exponentiation procedure in flavour space 
(see Appendix~A in Ref.~\cite{Bozzi:2005wk})
is completely analogous to the customary procedure that is usually applied to the scale evolution of the PDFs.

The main differences between the production of colourless systems and $t{\bar t}$ production start at the NLL level. In the case of $t{\bar t}$ production
the functions $g^{(2)}_N$, $g^{(3)}_N$ and so forth have an additional component 
\cite{Zhu:2012ts, Li:2013mia, Catani:2014qha}
due to soft-parton radiation at wide angles with respect to the direction of the colliding partons (hadrons). This component is produced by the non-vanishing colour charge of $t$ and ${\bar t}$ and, as recalled in Sect.~\ref{sec:intro},
it embodies the effect of soft radiation from the heavy-quark final state and from
initial-state and final-state interferences.
Soft wide-angle radiation leads to two main (and related) effects in the structure
of the resummation formulae in Eqs.~(\ref{eq:ff}), (\ref{eq:gexp}) and 
(\ref{eq:hfun}): dependence on the scattering angle and presence of colour correlations. The functions $g^{(k)}_N$ with $k\geq 2$ in Eq.~(\ref{eq:gexp})
acquire a dependence on the scattering angle $\theta$ (such dependence is instead absent in the production of colourless systems) and both these functions and the factor $\cal H$ in Eq.~(\ref{eq:ff}) are matrices in the colour space of the underlying LO production processes $q{\bar q} \to t{\bar t}$ and 
$gg \to t{\bar t}$ (we recall that the dependence on the colour indices is neglected in the notation of Eqs.~(\ref{eq:ff}) and (\ref{eq:gexp})).

The explicit expressions of the NLL+NLO contributions $g^{(2)}_N$ and ${\cal H}^{(1)}$
in Eqs.~(\ref{eq:gexp}) and (\ref{eq:hfun}) can be worked out from the results in Ref.~\cite{Catani:2014qha}. The term ${\cal H}^{(1)}$ (including its dependence on colour indices) is related to the hard-virtual amplitude ${\widetilde {\cal M}}^{(1)}$
in Eq.~(29) of Ref.~\cite{Catani:2014qha}. The soft wide-angle component of 
$g^{(2)}_N$ and, more generally, of ${\cal G}_N$ is related to the resummation factor
${\bf V}(b,M;y_{34})$ in Eqs.~(15) and (16) of Ref.~\cite{Catani:2014qha}. The rapidity variable
$y_{34}$ of Ref.~\cite{Catani:2014qha} is directly related to the scattering angle 
$\theta$ ($y_{34}= \ln[(1+\beta \cos\theta)/(1-\beta \cos\theta)] \,, \;
\beta= {\sqrt{1-4m_t^2/M^2}}$). The resummation factor ${\bf V}$ is the exponential
of a soft anomalous dimension matrix, ${\bf \Gamma}_t$, in colour space.
Therefore, the exponentiated form in Eq.~(\ref{eq:gexp}) requires a proper diagonalization procedure \cite{Kidonakis:1998nf}
of the soft anomalous dimension matrix with respect to its colour indices (such procedure is formally similar to the diagonalization with respect to flavour parton indices that we have previously mentioned). We have explicitly worked out the colour space diagonalization of the one-loop soft anomalous dimension 
${\bf \Gamma}_t^{(1)}$ (see\footnote{
A relative sign is mistyped in the right-hand side of the kinematical relation (34) in Ref.~\cite{Catani:2014qha}. The correct result is obtained by performing the replacement 
$\ln(m_T^2/m^2) \to - \ln(m_T^2/m^2)$ in the right-hand side of Eq.~(34) therein.}
Eq.~(33) in Ref.~\cite{Catani:2014qha}), whose eigenvalues contribute to the NLL function 
$g^{(2)}_N$  in Eq.~(\ref{eq:gexp}).
The factor ${\cal H}_N^{(0)}$ in Eq.~(\ref{eq:hfun}) is a colour space matrix and,
specifically, the colour space matrix elements of $d{\hat \sigma}_{c{\bar c}}^{(0)}{\cal H}_N^{(0)}$ are obtained by projecting the LO cross section onto the
eigenvectors of the soft anomalous dimension ${\bf \Gamma}_t^{(1)}$ 
(in the case of production of colourless systems we simply have ${\cal H}_N^{(0)}=1$).

We recall \cite{Bozzi:2005wk} that the resummed 
factor
$\exp\{{\cal G}(\as,\tL)\}$ of Eq.~(\ref{eq:ff}) is singular at very large
values of $b$. The singularity occurs in the region where
$b \gtap 1/\Lambda_{\rm QCD}$, $\Lambda_{\rm QCD}$ being the momentum scale
of the Landau pole of the perturbative running coupling $\as(\mu^2)$.
This singularity is the `perturbative' signal of the onset of non-perturbative
(NP) phenomena at very large values of $b$ 
(which practically affect the region of very small transverse momenta).
A simple and customary procedure to include NP
effects is as follows. The singular behaviour of the perturbative form factor
$\exp\{{\cal G}(\as,\tL)\}$ is removed by using a regularization
prescription
and the resummed expression in Eq.~(\ref{eq:ff}) is then multiplied
by a NP form factor and it is inserted as integrand of the $b$ space integral
in Eq.~(\ref{eq:resw}).
In the present work we use the so called `$b_*$ prescription'
of Ref.~\cite{Collins:va},
which is obtained by performing
the replacement
\begin{equation}
\label{bstar}
b^2 \to b_*^2 = b^2 \;b_{\rm lim}^2/( b^2 + b_{\rm lim}^2) 
\end{equation}
in the $b$ dependence of ${\cal G}(\as,\tL)$. The value of the parameter 
$b_{\rm lim}$ has to be large ($b_{\rm lim} M \sim b_{\rm lim} Q~\gg~1$) but smaller than the
value of $b$ at which the singularity of $\exp\{{\cal G}(\as,\tL)\}$ takes
place 
(note that the replacement in Eq.~(\ref{bstar}) has a negligible effect at
small and intermediate values of $b$ since 
$b_*^2 = b^2(1+{\cal O}(b^2Q^2/b_{\rm lim}^2Q^2)) \simeq b^2$
if $bQ~\ltap~1$).

\section{Results}
\label{sec:result}

In this Section we present the numerical results of our calculation and we compare them to ATLAS and CMS Run 1 data.
As stated in Sect.~\ref{sec:intro}, LHC measurements of the transverse-momentum distribution of the top-quark pair have been carried out in $pp$ collisions
at $\sqrt{s}=7$, 8 and 13 TeV 
\cite{Chatrchyan:2012saa}-\cite{Sirunyan:2018wem}.
The measurements at $\sqrt{s}=7$~TeV are based on data sets with relatively-small
integrated luminosity. A similar comment applies to the available measurements at 
$\sqrt{s}=13$~TeV, with the exception of CMS results \cite{Sirunyan:2018wem}
that have been presented only very recently.
In the following we limit ourselves to considering $pp$ collisions at $
\sqrt{s}=8$~TeV.
The experimental data that we consider in our comparison are those from
ATLAS \cite{Aaboud:2016iot}, which correspond to an integrated luminosity of $20.2~{\rm fb}^{-1}$ and refer to the dilepton decay channels of the $t{\bar t}$ pair,
and those from CMS \cite{Khachatryan:2015oqa}, which correspond to an integrated luminosity of $19.7~{\rm fb}^{-1}$ and refer to both the lepton+jets and dilepton channels.

The structure of the resummed cross section is illustrated in Sect.~\ref{sec:theo}.
Our resummed calculation is implemented in a numerical program that is 
obtained as an extension to heavy-quark production
of analogous codes for Higgs \cite{deFlorian:2012mx} and vector \cite{Catani:2015vma} boson production. We implement the resummed calculation at fixed values of $M$ and 
$\cos\theta$ (see Eq.~(\ref{eq:resw})).
We limit ourselves to presenting numerical results for the differential cross section
$d\sigma/dq_T$, which is obtained by numerical integration over $M$ and 
$\cos\theta$. The regular component of the cross section (see Eq.~(\ref{eq:singreg}))
is computed by considering the complete fixed-order result \cite{Bonciani:2015sha}
and subtracting the perturbative expansion of the resummed component of the cross section at the corresponding fixed order. 

We are going to present resummed results at NLL+NLO accuracy and, for the sake of comparison, we also present fixed-order results for the $q_T$ distribution
of the $t{\bar t}$ pair up to 
${\cal O}(\as^3)$ and ${\cal O}(\as^4)$. The fixed-order results for $d\sigma/dq_T$ up to 
${\cal O}(\as^3)$ and ${\cal O}(\as^4)$ contribute to the 
$t{\bar t}$ total cross section at NLO and NNLO, respectively. However, in the region where $q_T\neq 0$, they formally\footnote{In spite of the `effective' meaning of the results in different regions of $q_T$, we always use the default labels NLO and NNLO according to the perturbative order in which the results contribute to the total cross section.} correspond to `effective'
LO \cite{Mangano:1991jk} and NLO calculations \cite{Dittmaier:2007wz,Melnikov:2010iu}, respectively. 
Our results at ${\cal O}(\as^3)$ and ${\cal O}(\as^4)$
are obtained by using the calculation
and the numerical program of Ref.~\cite{Bonciani:2015sha}.
The $q_T$ cross section at ${\cal O}(\as^4)$ is evaluated by using
the {\sc Munich} code \cite{Kallweit:Munich}, which
is also at the heart of the {\sc Matrix} framework \cite{Grazzini:2017mhc}.
{\sc Munich} provides a fully automated implementation of the 
NLO dipole subtraction
formalism 
\cite{Catani:1996jh,Catani:2002hc}
as well as an interface to the one-loop generator {\sc OpenLoops} 
\cite{Cascioli:2011va} to obtain all the required (spin- and colour-correlated) 
tree-level and one-loop amplitudes.
For the evaluation of tensor integrals we rely on the 
\textsc{Collier} library \cite{Denner:2016kdg}, which is based on the 
Denner--Dittmaier reduction techniques \cite{Denner:2002ii} of tensor integrals
and on the scalar integrals of Ref.~\cite{Denner:2010tr}.
In {\sc OpenLoops}
problematic phase space points are addressed with a rescue system 
that uses the quadruple-precision implementation of the 
OPP method 
in \textsc{CutTools}~\cite{Ossola:2007ax}
with scalar integrals from \textsc{OneLOop}~\cite{vanHameren:2010cp}.

To evaluate the normalized $q_T$ distribution (see Eq.~(\ref{eq:norm_dist}) below) at NNLO,
the NNLO total cross section is required. 
We compute it by using
the numerical program {\sc Top++}
\cite{Czakon:2011xx}, which implements the NNLO calculation of 
Ref.~\cite{Baernreuther:2012ws}.

In the following we first focus our discussion on the $q_T$ distribution of the
$t{\bar t}$ pair
at fixed order (Sect.~\ref{sec:fo}) and then we move to present our resummed results (Sect.~\ref{sec:res}).

\subsection{Fixed-order results}
\label{sec:fo}

We start the presentation of our numerical results by considering QCD calculations at fixed order.
To compute the NLO and NNLO hadronic $q_T$ cross section 
(see Eq.~(\ref{eq:triplesigma}))
we use the NNPDF3.0~\cite{Ball:2014uwa} sets of PDFs at NLO and NNLO with $\alpha_{\mathrm{S}}(m_Z)=0.118$. Correspondingly, the scale ($\mu$) dependence of the strong coupling constant 
$\alpha_{\mathrm{S}}(\mu^2)$ is evaluated at two and three loop accuracy
in the NLO and NNLO calculations, respectively.
The value of the pole mass of the top quark is $m_t=173.3$~GeV.
As for the factorization ($\mu_F$) and renormalization ($\mu_R$) scales, we choose $\mu_F=\mu_R=m_t$ as central value, and we consider variations of $\mu_F$ and 
$\mu_R$ around this central value.

Using these parameters the values of the $t\bar{t}$ total cross section at central scales are $\sigma=224.1$~pb at NLO and $\sigma=243.5$~pb at NNLO.
We note that the NNLO corrections increase the NLO total cross section by approximately 9\%.

\begin{figure}[th]
\centering
\includegraphics[width=0.58\textwidth,angle=90]{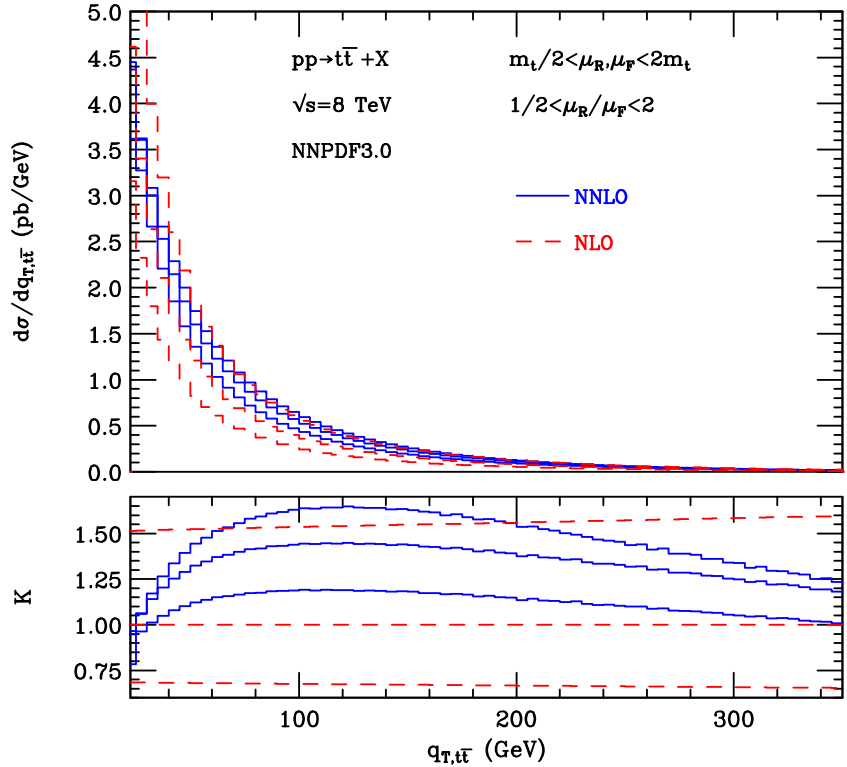}\hfill
\caption{\label{fig:NLO_LO}
{\em The $q_T$ cross section $d\sigma/dq_T$ of the $t\bar{t}$ pair produced in $pp$ collisions at
${\sqrt s}=8$~TeV: 
NLO (red dashed) and NNLO (blue solid) theoretical predictions at central scales and
including scale dependence. The lower panel shows the ratio $K$ of the NNLO and NLO results (blue solid), and the relative scale dependence at NLO (red dashed).}}
\end{figure}

In Fig.~\ref{fig:NLO_LO} we consider the region where $q_T > 20$~GeV and we present
the differential cross section $d\sigma/dq_T$ of the 
$t\bar{t}$ pair at the LHC Run I ($\sqrt{s}=8\,\mathrm{TeV}$) at NLO and NNLO accuracy. We present results at central values of the scales and including scale
variation effects.
The bands are obtained by independently varying the factorization  and renormalization scales by a factor ot two around the central values 
$\mu_F=\mu_R=m_t$ (i.e., we consider the range 
$0.5m_t\leq \{\mu_F,\,\mu_R\}\leq 2m_t$), with the constraint $0.5\leq\mu_F / \mu_R\leq 2$. At the NLO the scale dependence is at the level of roughly $\pm 40\,\%$. At NNLO the scale dependence is relatively flat in the intermediate region 80~GeV$\leq q_T \leq 200$~GeV and 
it is about $\pm 18\,\%$.
The NNLO scale dependence slightly decreases at larger values of $q_T$ and it also
decreases at smaller values of $q_T$.

The lower panel of Fig.~\ref{fig:NLO_LO} shows the $K$ factor, which is obtained by normalizing the NNLO band with respect to the NLO result at $\mu_F=\mu_R=m_t$. 
Using the same normalization we also show the scale dependence of the NLO result.
At central values of the scales the impact of the NNLO corrections ranges from about $-5\,\%$ at $q_T\sim 20$ GeV to about $+18\,\%$ at $q_T\sim 350$ GeV. The $K$ factor is relatively flat and larger in the intermediate $q_T$ region (80~GeV$\leq q_T \leq 160$~GeV) where its value is about 1.4\,.

As shown in Fig.~\ref{fig:NLO_LO}, at intermediate and large values of $q_T$ the scale variation bands of the NLO and NNLO results overlap and the NNLO result has a reduced scale dependence. In this $q_T$ region the NNLO scale dependence can consistently be used as an approximate estimate of the theoretical uncertainty of the NNLO prediction. At smaller values of $q_T$ (say, $q_T \ltap 60$~GeV), the NNLO scale dependence strongly decreases by decreasing $q_T$ and the size of the $K$ factor
becomes close to unity at $q_T \sim 30$~GeV. We anticipate
(see the discussion of the results in Fig.~\ref{fig:NLO_data}-right)
that the behaviour of the NNLO radiative corrections at $q_T \sim 30$~GeV
should not be regarded as a signal of perturbative convergence: in constrast,
it is just a consequence and an artifact of the order-by-order perturbative instability of the shape of the $q_T$ cross section in the small-$q_T$ region.

\begin{figure}[th]
\begin{center}
\includegraphics[width=0.58\textwidth,angle=90]{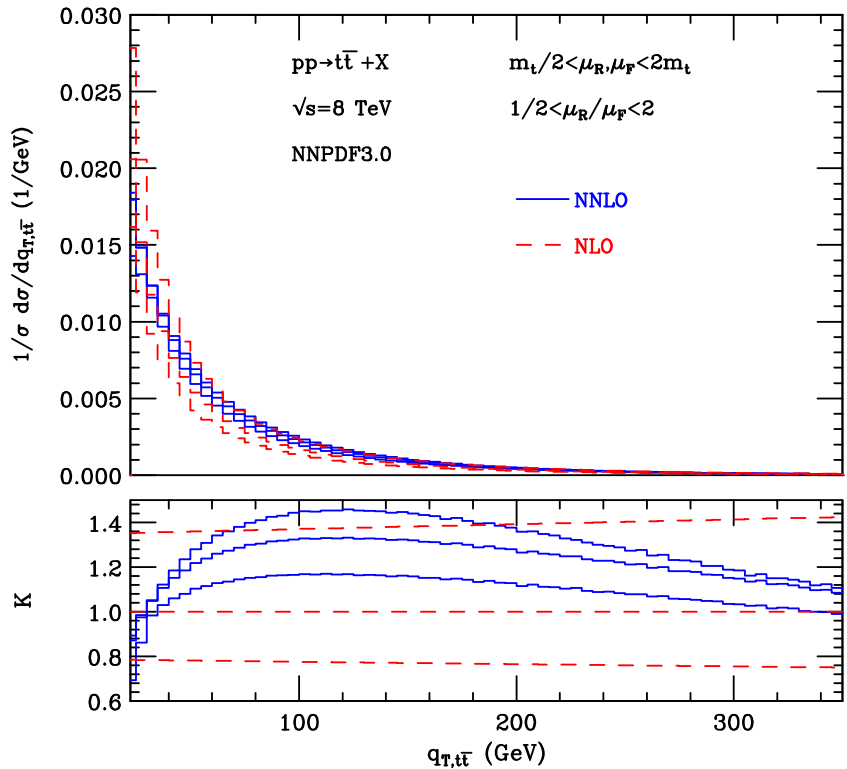}\hfill
\caption{\label{fig:NLO_LO_norm}
{\em Same as in Fig.~\ref{fig:NLO_LO} for the normalized $q_T$ distribution
$1/\sigma\,(d\sigma/dq_T)$.
}}
\end{center}
\end{figure}

The results in Fig.~\ref{fig:NLO_LO_norm} are analogous to those in 
Fig.~\ref{fig:NLO_LO}, but they refer to the normalized $q_T$ spectrum of the $t\bar{t}$ pair. We consider 
the normalized $q_T$ distribution, $\frac{1}{\sigma}\frac{d\sigma}{dq_T}$, of $t\bar{t}$ pairs at the LHC Run I ($\sqrt{s}=8\,\mathrm{TeV}$) at NLO and NNLO accuracy. More precisely,
we define
\begin{equation}
\left(\frac{1}{\sigma}\frac{d\sigma}{dq_T}\right)_{(\mathrm{N})\mathrm{NLO}}(\mu_F,\mu_R)\equiv\frac{1}{\sigma_{(\mathrm{N})\mathrm{NLO}}(\mu_F,\mu_R)}\left(\frac{d\sigma}{dq_T}\right)_{(\mathrm{N})\mathrm{NLO}}(\mu_F,\mu_R)\,,
\label{eq:norm_dist}
\end{equation}
and the two factors, $1/\sigma$ and $d\sigma/dq_T$, on the right-hand side of Eq.~(\ref{eq:norm_dist}) are evaluated by using the same PDFs and the same values of the renormalization and factorization scales. 
The scale variation bands and the $K$ factor are obtained as 
in the case of the cross section results of Fig.~\ref{fig:NLO_LO}.
%

From the results in Fig.~\ref{fig:NLO_LO_norm}, we see that the NLO scale dependence of the normalized distribution is at the level of roughly $\pm 30\,\%$.
At NNLO the scale dependence is about $\pm 12\,\%$ in the region where
80~GeV$\leq q_T \leq 200$~GeV.
At central values of the scales the impact of the NNLO corrections
(lower panel in Fig.~\ref{fig:NLO_LO_norm})
ranges from about $-10\,\%$ at $q_T\sim 20$~GeV to about $+10\,\%$ at 
$q_T\sim 350$~GeV.
The $K$ factor is relatively flat and larger in the intermediate $q_T$ region 
(80~GeV$\leq q_T \leq 160$~GeV) where its value is about 1.3\,.

Comparing the results in Fig.~\ref{fig:NLO_LO} with those in 
Fig.~\ref{fig:NLO_LO_norm}, we can see that they have very similar features. The main
differences are that the results for the normalized $q_T$ distribution have a decreased scale dependence (at both NLO and NNLO) and a smaller NNLO $K$ factor
in the intermediate $q_T$ region. 
This implies that QCD radiative corrections to $d\sigma/dq_T$ include contributions
with a small dependence on $q_T$ that do not affect the shape of the $q_T$
spectrum. Their effect and the related scale dependence partly cancel in the ratio
between $d\sigma/dq_T$ and the total cross section $\sigma$.

In Fig.~\ref{fig:NLO_data} we compare the fixed-order predictions for the normalized $q_T$ distribution with the experimental data from ATLAS \cite{Aaboud:2016iot} and CMS~\cite{Khachatryan:2015oqa}. The plots in Fig.~\ref{fig:NLO_data} show both the NLO and NNLO predictions with their scale uncertainties normalised to the NNLO result at central values of the scales, and the
relative deviation of the data from such central NNLO prediction.
Thus the results presented in Fig.~\ref{fig:NLO_data} refer to the fractional differences (X-`theory')/`theory', where X=\{NLO, NNLO, data\} and the reference theoretical result (`theory') is the NNLO prediction at central scales.
We point out that the theoretical predictions are obtained by using $q_T$ bins
with a constant size of 5~GeV,
while the comparison to the data is done by using exactly the same bin sizes that are used in the experimental measurements. The two panels in Fig.~\ref{fig:NLO_data}
present the same content by using either a linear (left panel) or 
a logarithmic (right panel) $q_T$ scale on the horizontal axis. 
The region where $q_T <20$~GeV ($q_T > 150$~GeV) is excluded 
in the left (right) panel.

\begin{figure}[th]
\centering
\hspace*{-0.3cm}
\subfigure[]{
\includegraphics[width=0.3\textwidth,angle=90]{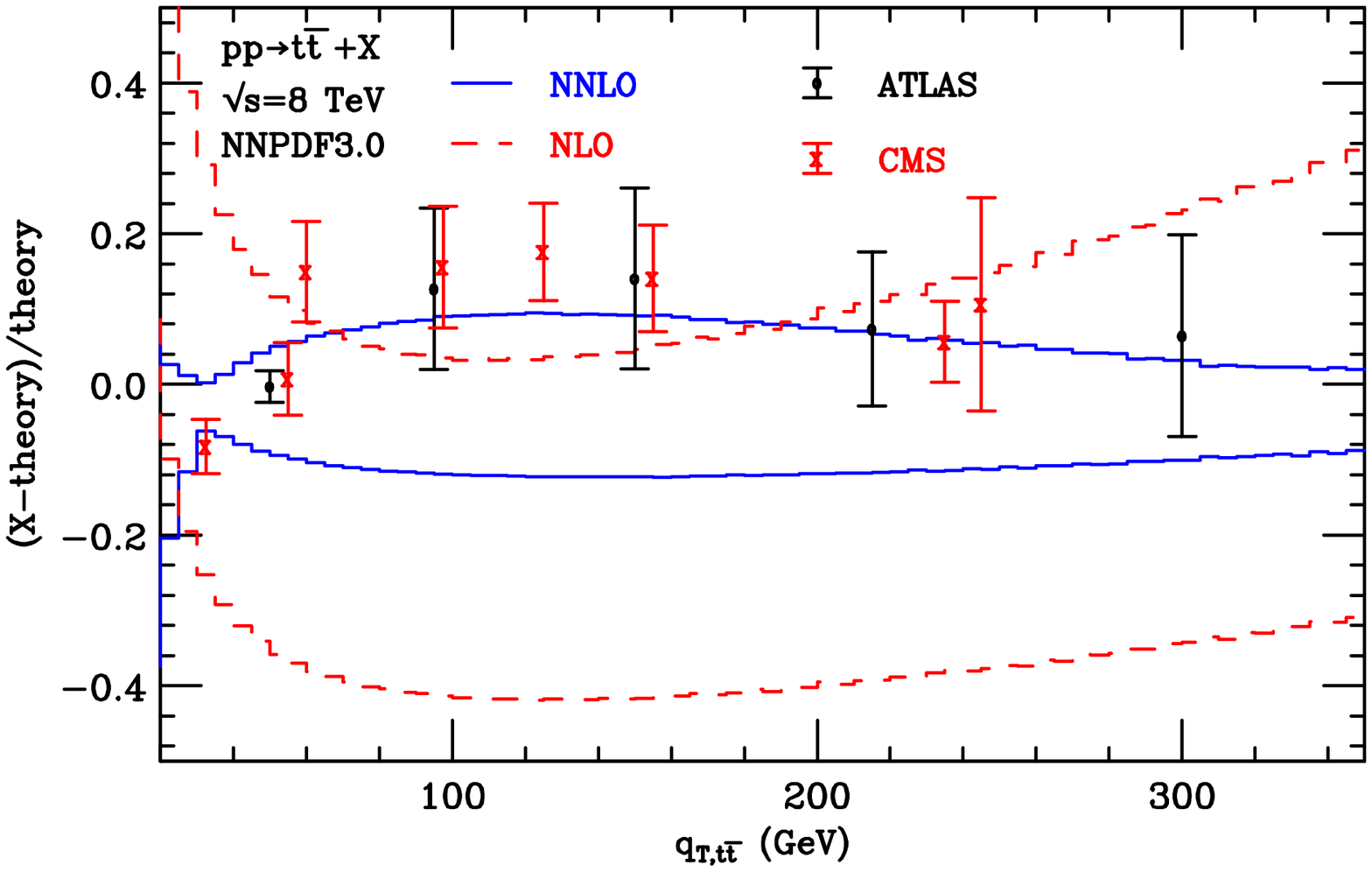}
}
\subfigure[]{
\includegraphics[width=0.3\textwidth,angle=90]{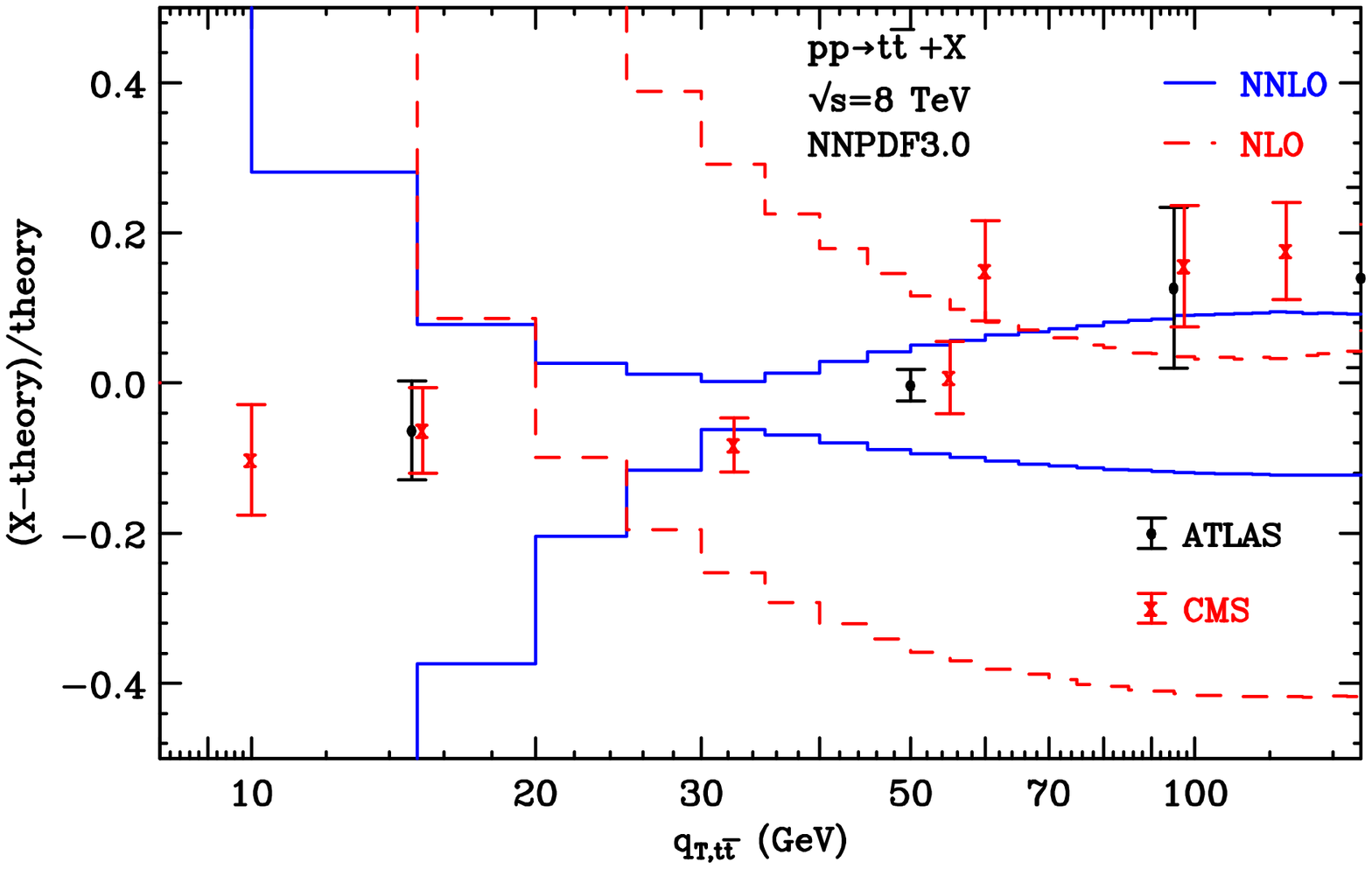}
}
\caption{\label{fig:NLO_data}
{\em Fractional difference of NLO predictions (red dashed), NNLO predictions
(blue solid) and LHC data \cite{Aaboud:2016iot,Khachatryan:2015oqa} 
with respect to the NNLO result at $\mu_F=\mu_R=m_t$.
The results refer to the normalized $q_T$ distribution $1/\sigma\,(d\sigma/dq_T)$
of $t{\bar t}$ pairs at ${\sqrt s}=8$~TeV.
The two panels highlight (a) the intermediate and large $q_T$ region and 
(b) the small-$q_T$ region.}}
\end{figure}

We start our discussion of the results in Fig.~\ref{fig:NLO_data} by considering the
region of intermediate and high values of $q_T$ (left panel).
Figure~\ref{fig:NLO_data}-left and
the lower panel of Fig.~\ref{fig:NLO_LO_norm} present the same fixed-order results, but the role of the NNLO and NLO central values as reference theoretical result is exchanged between the two figures, and the overall shapes of the NLO and NNLO bands are exchanged accordingly. At intermediate and large values of $q_T$ we see that
the data are consistent with the NNLO prediction within uncertainties. Only two data points overshoot the NNLO prediction, but the difference is 
smaller than one standard deviation.
We also note that the experimental uncertainties and the NNLO uncertainties, estimated through scale variations, are of similar size.
The data and the NLO predictions are not inconsistent within uncertainties.
The NNLO corrections improve the agreement with the data by both increasing the central value of the theoretical prediction and decreasing its scale uncertainty.

We now move to consider the small-$q_T$ region (Fig.~\ref{fig:NLO_data}-right). 
We first discuss the behaviour of the fixed-order results.
Starting from $q_T \sim 60$~GeV and decreasing the value of $q_T$,
the NNLO scale dependent band shrinks, by reaching a minimum size around 
$q_T \sim 30$~GeV, where the ratio NLO/NNLO at central scales is close to unity.
Then, by further decreasing $q_T$, the size of the NNLO band 
quickly increases and the NNLO scale dependence is of the order of $\pm 50\%$
or larger at $q_T\sim 10$~GeV. Moreover, in the region below $q_T \sim 20$~GeV, the NLO band quickly departs from the NNLO band and the ratio NLO/NNLO at central scales
suddenly becomes much different from unity: in this $q_T$ region the fixed-order expansion cannot reliably predict the detailed shape of the $q_T$ spectrum.
At $q_T \sim 30$~GeV, the fixed-order expansion is apparently quite stable but this stability has to be regarded as partly accidental since it takes place very close to the low-$q_T$ region where the convergence of the fixed-order expansion is spoiled. 

The behaviour of the NLO and NNLO results at small $q_T$ is produced by the large logarithmic terms that we have discussed in Sects.~\ref{sec:intro} and \ref{sec:theo}.
The qualitative features of the NLO and NNLO results in Fig.~\ref{fig:NLO_data}-right
are indeed completely similar to those of analogous fixed-order results for other processes that are affected by large logarithmic contributions in the small-$q_T$ region (see, e.g., Fig.~3 in Ref.~\cite{Bozzi:2008bb} and accompanying comments for a related discussion in the context of vector boson production). 

In the small-$q_T$ region the perturbative resummation of the large logarithmic terms is necessary to reliably predict the detailed shape of the $q_T$ spectrum of the 
$t{\bar t}$ pair. Nonetheless, we note 
that the data points in Fig.~\ref{fig:NLO_data}-right are still perfectly consistent with the NNLO result within uncertainties.
This is a consequence of the fact that the experimental measurements use 
relatively-large bin sizes in $q_T$: the first bin covers the region up to 
$q_T=20$~GeV for the CMS lepton+{\rm jets} data set 
and up to $q_T=30$~GeV for the other two measurements
(in Fig.~\ref{fig:NLO_data} each data point is placed at the midpoint of the corresponding $q_T$ bin) . Using such relatively-large bin sizes, the singular behavior of the fixed-order calculation at $q_T\to 0$ is smeared out, 
and it turns out that 
at NNLO a sensible central result\footnote{We remark that in the lowest-$q_T$ bin (which includes $q_T=0$) the NNLO result for $1/\sigma (d\sigma/dq_T)$
includes the complete contributions at the first three perturbative orders
(i.e., ${\cal O}(\as^2+\as^3+\as^4)$) in the computation of both
$d\sigma/dq_T$ and $\sigma$.}
is obtained, though it is affected by very large scale uncertainties. 

In our comments throughout this subsection we have made distinctions among small, intermediate and large values of $q_T$. We notice that the $q_T$ spectrum of 
$t{\bar t}$ pairs at the LHC is quite broad and it has an average transverse momentum
$\langle q_T \rangle$ of approximately 50~GeV. At the NLO with central value of the scales we have $\langle q_T \rangle =50.1$~GeV. We note that such value of  
$\langle q_T \rangle$ is roughly three times larger than the NLO value 
\cite{Frixione:1995fj}
for $t{\bar t}$ production at the Tevatron ($p{\bar p}$ collisions at 
${\sqrt s}=1.8$~TeV).
The value of $\langle q_T \rangle$ is approximately given by the proportionality relation $\langle q_T \rangle \sim C \,\as \,m_t$, where the proportionality factor $C$ depends on the underlying QCD dynamics and it has a weak dependence on ${\sqrt s}$
($C$ slowly increases by increasing ${\sqrt s}$ because of the larger available phase space). The contribution to $C$ from initial-state radiation tends to be proportional to the colour coefficient $C_A=N_c=3 \;(C_F=(N_c^2 -1)/(2N_c)=4/3)$ for production subprocesses that are due to $gg$ fusion ($q{\bar q}$ annihilation). This colour 
coefficient dependence qualitatively explains why the $q_T$ spectrum at the LHC is much broader than the spectrum at the Tevatron. Indeed, the $t{\bar t}$ pair is mostly produced by $gg$ fusion at the LHC, whereas $q{\bar q}$ annihilation
dominates at the Tevatron (this is a consequence of the relative differences between 
$gg$ and $q{\bar q}$ PDF luminosities in $pp$ collisions at the LHC and 
$p{\bar p}$ collisions at the Tevatron).

\subsection{Resummed results}
\label{sec:res}

In the following we present our resummed results at NLL+NLO accuracy and we compare them
with the LHC data. To compute the resummed cross sections, we use the NNPDF3.0 NLO
PDFs \cite{Ball:2014uwa},
with the scale dependence of $\alpha_{\mathrm{S}}(\mu^2)$ evaluated at two-loop order. The pole mass of the top quark is $m_t=173.3$~GeV as in our fixed-order calculations.
As discussed in Sect.~\ref{sec:theo}, the resummed predictions depend on renormalization, factorization and resummation scales. 
The effect of factorization and renormalization scale variations is computed as in the fixed-order calculations of Sect.~\ref{sec:fo} by using $\mu_F=\mu_R=m_t$ as central
value of these scales.
We choose $Q=m_t$ as central value of the resummation scale $Q$, and we consider resummation scale variations in the range $m_t/2 < Q < 2m_t$.
The parameter $b_{\rm lim}$ in Eq.~(\ref{bstar}) is set to the value
$b_{\rm lim}=3~{\rm GeV}^{-1}$.

We note that the auxiliary scales $\mu_F$, $\mu_R$ and $Q$ have to be chosen of the order of the typical hard scale
of the cross section to avoid a parametrically-large scale dependence from missing higher-order contributions (in the context of both fixed-order and resummed perturbation theory).
In the case of the $q_T$ spectrum at fixed invariant mass $M$ of the $t{\bar t}$ pair (Sect.~\ref{sec:theo}), the typical hard scale of the cross section is $M$.
In the case of the $q_T$ spectrum integrated over $M$ (which is considered in all the results of Sect.~\ref{sec:result}), the typical hard scale turns out to be of the order of $m_t$
since the bulk of the invariant mass distribution is concentrated within a narrow region where $M\sim 2m_t$ (see e.g. Fig.~8 in the second paper of Ref.~\cite{Czakon:2016ckf}).

\begin{figure}[th]
\centering
\includegraphics[width=0.58\textwidth,angle=90]{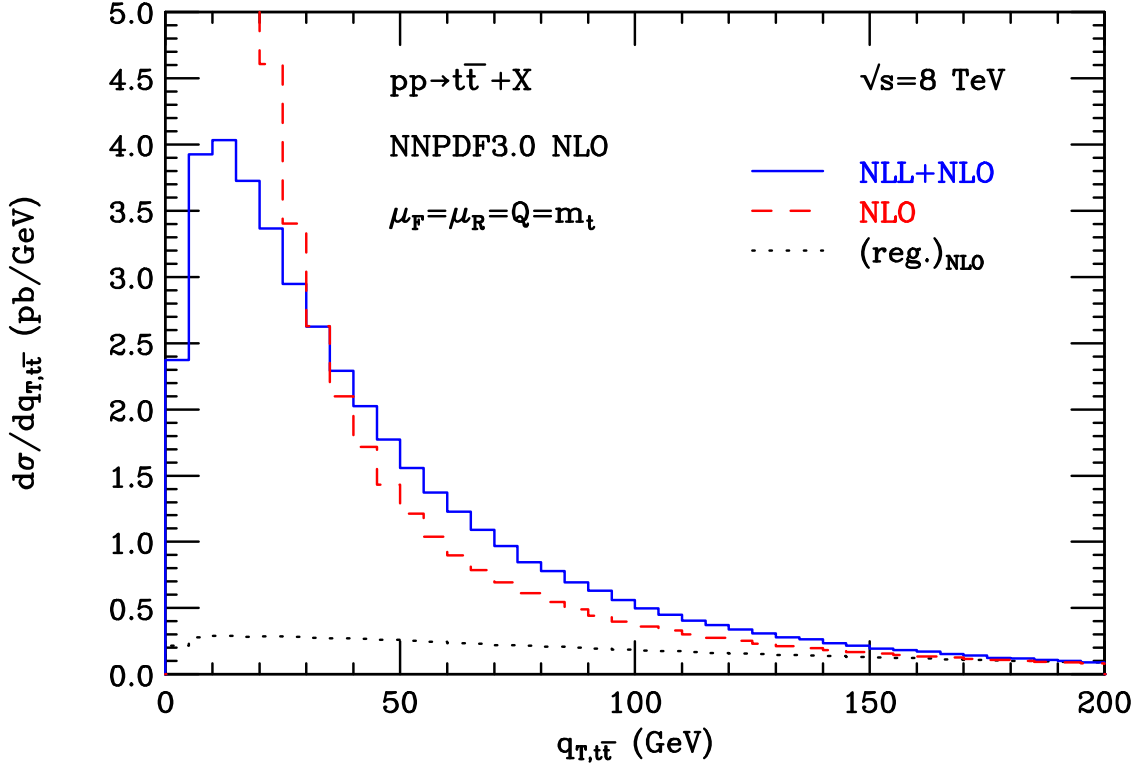}\hfill
\caption{\label{fig:NLL}
{\em 
The transverse-momentum cross section $d\sigma/dq_T$ of the $t\bar{t}$ pair at the LHC ($\sqrt{s}=8~\mathrm{TeV}$) computed through resummation at NLL+NLO accuracy (blue solid). The resummed result at central scales
($\mu_R=\mu_F=Q=m_t$) is compared to the corresponding NLO result 
(red dashed). The contribution of the regular component (black dotted) to the NLL+NLO result is also shown.}}
\end{figure}

In Fig.~\ref{fig:NLL} 
the NLL+NLO 
cross section $d\sigma/dq_T$
(solid line) at
the central scales ($\mu_F=\mu_R=Q=m_t$) is compared with the
corresponding NLO result (dashed line) and with the regular component 
$d\sigma^{(\rm reg)}/dq_T$ (see Eq.~(\ref{eq:singreg}))
of the cross section (dotted line).
At small values of $q_T$ the NLO result becomes arbitrarily large by decreasing 
$q_T$ towards lower (non-vanishing) values. 
In the first $q_T$-bin of Fig.~\ref{fig:NLL} ($q_T \leq 5$~GeV), which includes 
$q_T=0$, the NLO result is negative (the negative value is outside the vertical scale of the plot in Fig.~\ref{fig:NLL}). Such behaviour of the NLO result is definitely unphysical.
The resummation of the small-$q_T$ logarithms
leads to a physically well-behaved distribution at small transverse momenta, with a kinematical\footnote{After resummation $d\sigma/dq_T^2 \to {\rm const.}$ as $q_T\to 0$. Therefore, $d\sigma/dq_T=2q_T \,d\sigma/dq_T^2$ vanishes as $q_T\to 0$ and it has a peak due to the kinematical (Jacobian) factor of $2q_T$.} 
peak in the region where 
$q_T\sim 10~\mathrm{GeV}-15~\mathrm{GeV}$.
At large values of $q_T$, the NLL+NLO result tends to the corresponding NLO result. 
In the small-$q_T$ region the NLL+NLO result is dominated by resummation, although the contribution of the regular component is not negligible 
(it is approximately $9\,\%$).
In the region of intermediate values of $q_T$ (say, around $100$~GeV), the contribution of the regular component increases to about $ 35\,\%$
of the NLL+NLO result. At larger values of $q_T$ the contribution of the regular component
sizeably increases, indicating that  
the logarithmic terms are no longer dominant and that the resummed
calculation cannot improve upon the predictivity of the fixed-order expansion.

\begin{figure}[th]
\centering
\subfigure[]{
\includegraphics[width=0.43\textwidth,angle=90]{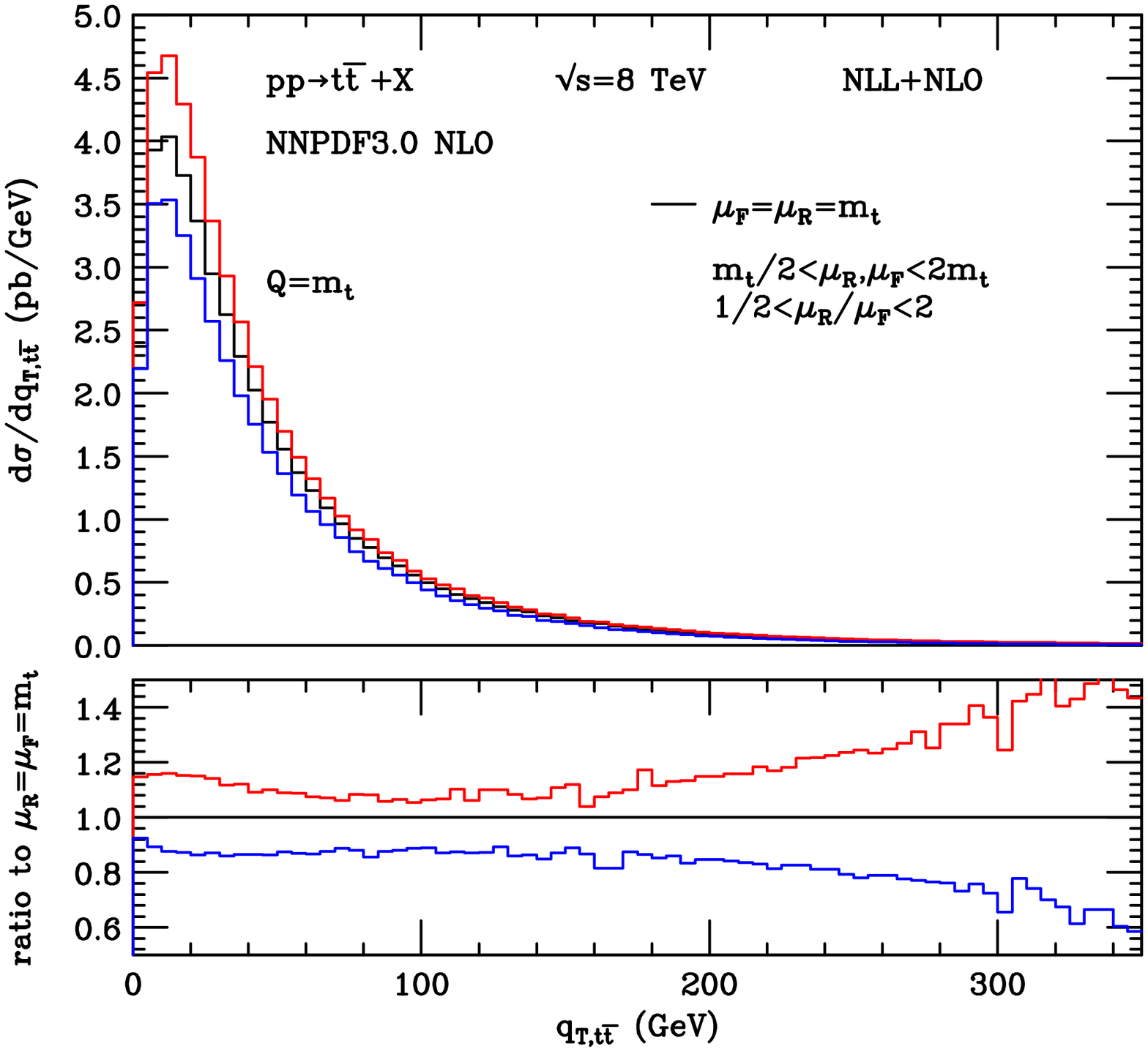}
}
\subfigure[]{
\includegraphics[width=0.43\textwidth,angle=90]{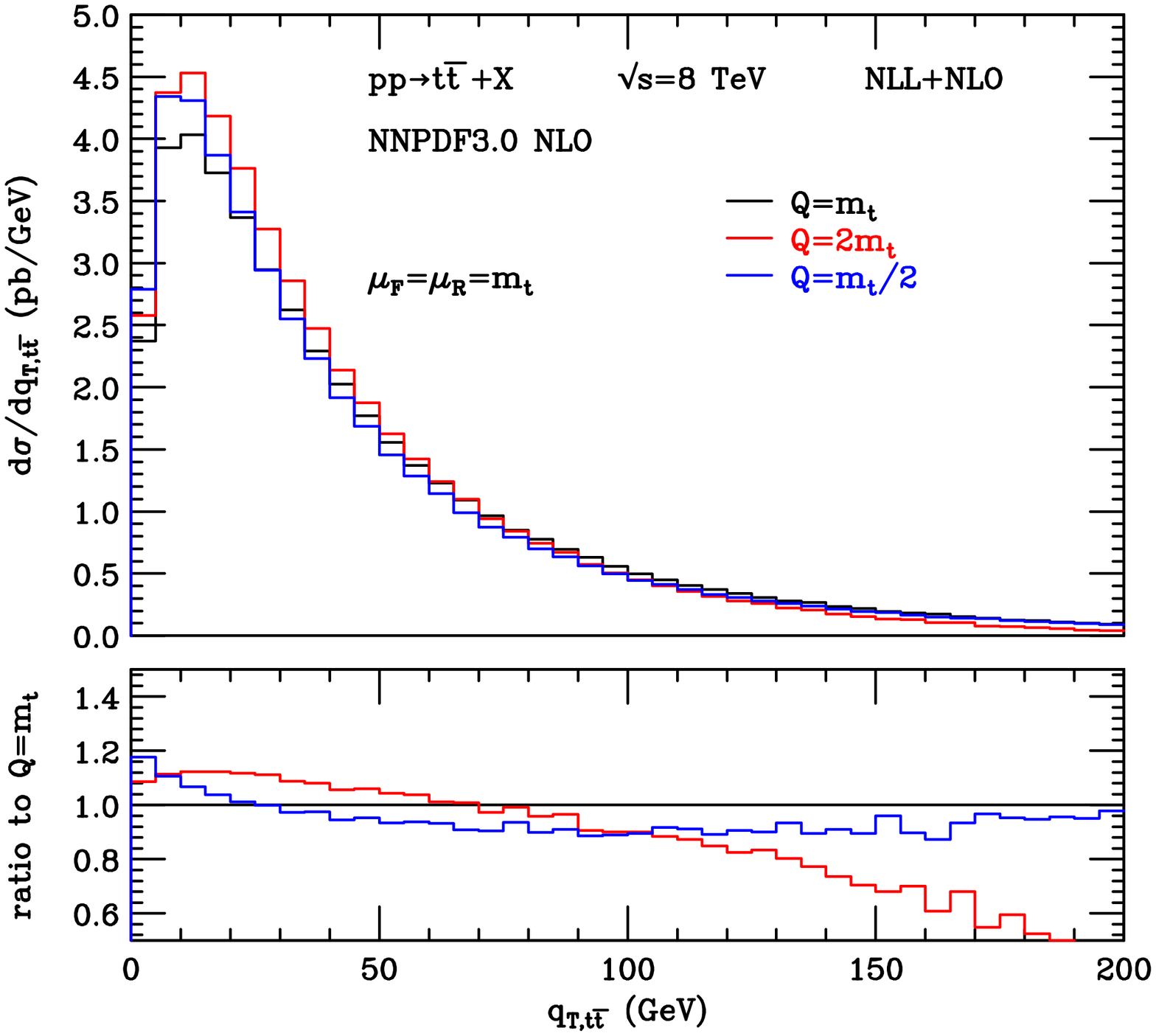}
}
\caption{\label{fig:nll_mur_muf}
{\em 
The transverse-momentum cross section $d\sigma/dq_T$ of the $t\bar{t}$ pair at the LHC ($\sqrt{s}=8~\mathrm{TeV}$) computed at NLL+NLO accuracy. The bands (blue and red lines) are obtained by varying $\mu_F$ and $\mu_R$ (left) and $Q$ (right) as described in the text. The lower panels present the scale variation bands relative to the result at central scales ($\mu_R=\mu_F=Q=m_t$).}}
\end{figure}

In Fig.~\ref{fig:nll_mur_muf} we 
show the scale dependence of our resummed results. 

In Fig.~\ref{fig:nll_mur_muf}-left 
we consider the effect of variations of the renormalization and factorization scales 
by keeping the resummation scale fixed at the central value $Q=m_t$.
The bands are obtained by independently varying $\mu_R$ and $\mu_F$ as usually done throughout this paper.
The scale variation in the peak region is at the level of about $\pm 20$\%. At intermediate values of $q_T$ the scale variation band shrinks a bit, while in the region of large values of $q_T$ the scale dependence increases dramatically, reaching even the level of $\pm 100$\%.

In Fig.~\ref{fig:nll_mur_muf}-right
we consider the effect of resummation scale variations. 
The results
are obtained
by fixing $\mu_R=\mu_F=m_t$ and considering three values 
$Q=\{m_t/2, m_t, 2m_t\}$ of the resummation scale $Q$.
Performing variations of the resummation scale,
we can get further insight on the size of yet uncalculated 
higher-order logarithmic contributions at small and intermediate values of 
$q_T$.
In the peak region
we find that the 
resummation scale dependence is about $15$\% and positive.
Resummation scale effects of similar size are found at intermediate values of $q_T$, while at very large transverse momenta the resummation scale dependence is negative and very large.

As recalled in Sect.~\ref{sec:theo}, our resummation procedure for the $q_T$ cross section formally reproduces the fixed-order result for the total cross section.
We find that
the integral over $q_T$ of our resummed NLL+NLO result of the $q_T$
spectrum is in agreement (for any values 
of $\mu_R, \mu_F$ and $Q$) with the value of
the NLO total cross section to better than 1\%,
thus checking
the numerical accuracy of the code. For example, at central values of the scales, 
the value of our total cross section at NLL+NLO accuracy is 223.8~pb, which is in excellent agreement with the value 224.1~pb of the NLO result.
We also notice that the average transverse momentum $\langle q_T \rangle$ is little affected by the resummation procedure. At central scales, using the resummed result
at NLL+NLO accuracy we obtain $\langle q_T \rangle=51.3$~GeV, which is very similar
to the NLO value 50.1~GeV.

At large values of $q_T$\footnote{We notice that the large-$q_T$ region gives a very small contribution to the total cross section.} the resummation of the logarithmic terms $\ln(q_T/M)$ is theoretically unjustified, since these terms are not the dominant radiative corrections in the large-$q_T$ region. Indeed, as previously observed in our comments on the results in Fig.~\ref{fig:NLL}, the fixed-order contribution of the regular component $d\sigma^{(\rm reg)}/dq_T$ to the $q_T$ cross section is sizeable at large values of $q_T$. The scale dependence of the NLL+NLO result is larger than the scale dependence of the fixed-order result (though the two results are consistent within scale uncertainties) at large $q_T$, and this fact further indicates that resummation 
is less predictive than the fixed-order expansion.
In the large-$q_T$ region the resummed prediction can simply be replaced by fixed-order predictions. Alternatively, a `smooth switching procedure'
between the resummed and fixed-order results 
(see, e.g., Refs.~\cite{deFlorian:2012mx} and \cite{Catani:2015vma})
can consistently be implemented at large values of $q_T$.

The NLL+NLO result has (at the formal level) a uniform theoretical accuracy 
throughout the region from small to intermediate values of $q_T$.
As discussed in Ref.~\cite{Bozzi:2005wk},
this is the consequence of the consistent combination (matching procedure) of the fixed-order term $d\sigma^{\rm (reg)}$ with the resummed contribution to 
$d\sigma^{\rm (sing)}$ (see Eq.~(\ref{eq:singreg})) and of the fact that the resummed calculation returns the fixed-order value of the total cross section after integration over $q_T$.

\begin{figure}[th]
\centering
\includegraphics[width=0.58\textwidth,angle=90]{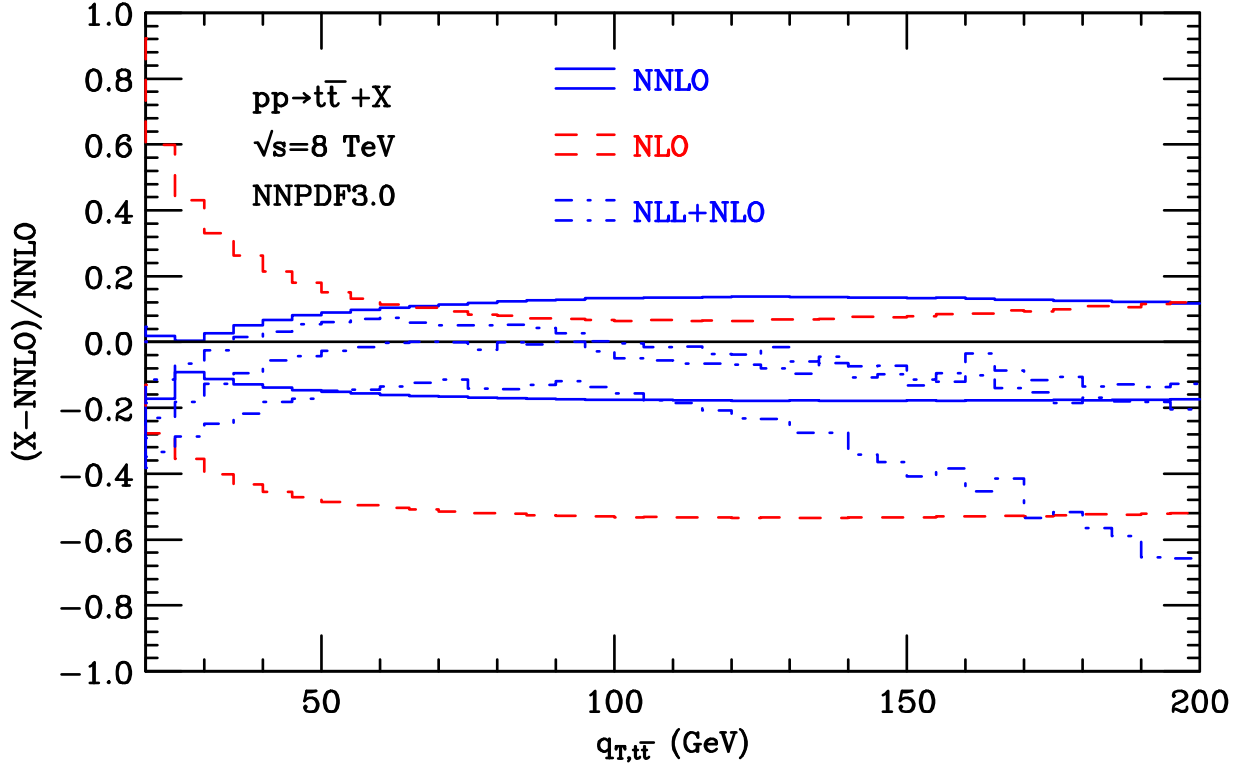}
\caption{\label{fig:X-NLO}
{\em Fractional difference of the scale dependent NLO (red dashed), NNLO (blue solid) and NLL+NLO (blue dot-dashed) results with respect to the NNLO result at $\mu_F=\mu_R=m_t$.
The results refer to the differential cross section $d\sigma/dq_T$
of $t{\bar t}$ pairs at ${\sqrt s}=8$~TeV.
}}
\end{figure}

A direct quantitative comparison
between fixed-order and resummed results at intermediate values of $q_T$ is presented in Fig.~\ref{fig:X-NLO}. We consider $d\sigma/dq_T$ and in Fig.~\ref{fig:X-NLO}
we present the fractional difference of the scale dependent NLO (dashed), NNLO (solid) and NLL+NLO (dot dashed) results with respect to the NNLO result at central
scales ($\mu_F=\mu_R=m_t$). 
The scale dependence of the NLO and NNLO results is obtained by the seven-point scale variation of $\mu_R$ and $\mu_F$. Therefore,
the NNLO and NLO bands in Fig.~\ref{fig:X-NLO} exactly corresponds to the bands in Fig.~\ref{fig:NLO_LO}
(though the role of the NNLO and NLO central values as reference theoretical result is exchanged between the two figures). 
The scale dependence of the NLL+NLO result is obtained by taking the envelope of the seven-point scale variation of $\mu_R$ and $\mu_F$ at fixed resummation scale 
$Q=m_t$ (see Fig.~\ref{fig:nll_mur_muf}-left) and the variation of $Q$ by a factor of two at fixed $\mu_R=\mu_F=m_t$ (see Fig.~\ref{fig:nll_mur_muf}-right).

In Fig.~\ref{fig:X-NLO} we can see that the scale dependence of the NLL+NLO result
has a moderate size at intermediate values of $q_T$.
Moreover, in the region where 50~GeV$\ltap q_T\ltap 150$~GeV 
the NLL+NLO and NNLO central values are quite close, their difference being always smaller than $10\%$, and they have comparable scale dependence.
We conclude that the NLL+NLO and NNLO results are fully consistent in this 
intermediate region of transverse momenta.
The NLL+NLO calculation provides us with a QCD prediction that can be extended down to lower values of $q_T$ with a relatively-small perturbative uncertainty.
In particular, in the region where 20~GeV$\ltap q_T\ltap 50$~GeV  the NLL+NLO scale dependence is approximately constant and the central value of the NNLO result
tends to deviate from the NNLO result. This behaviour is a further indication that the reduction of the NNLO scale dependence at $q_T \sim 30$~GeV is accidental and it does underestimate the theoretical uncertainty of the NNLO result.

\begin{figure}[th]
\centering
\subfigure[]{
\includegraphics[width=0.3\textwidth,angle=90]{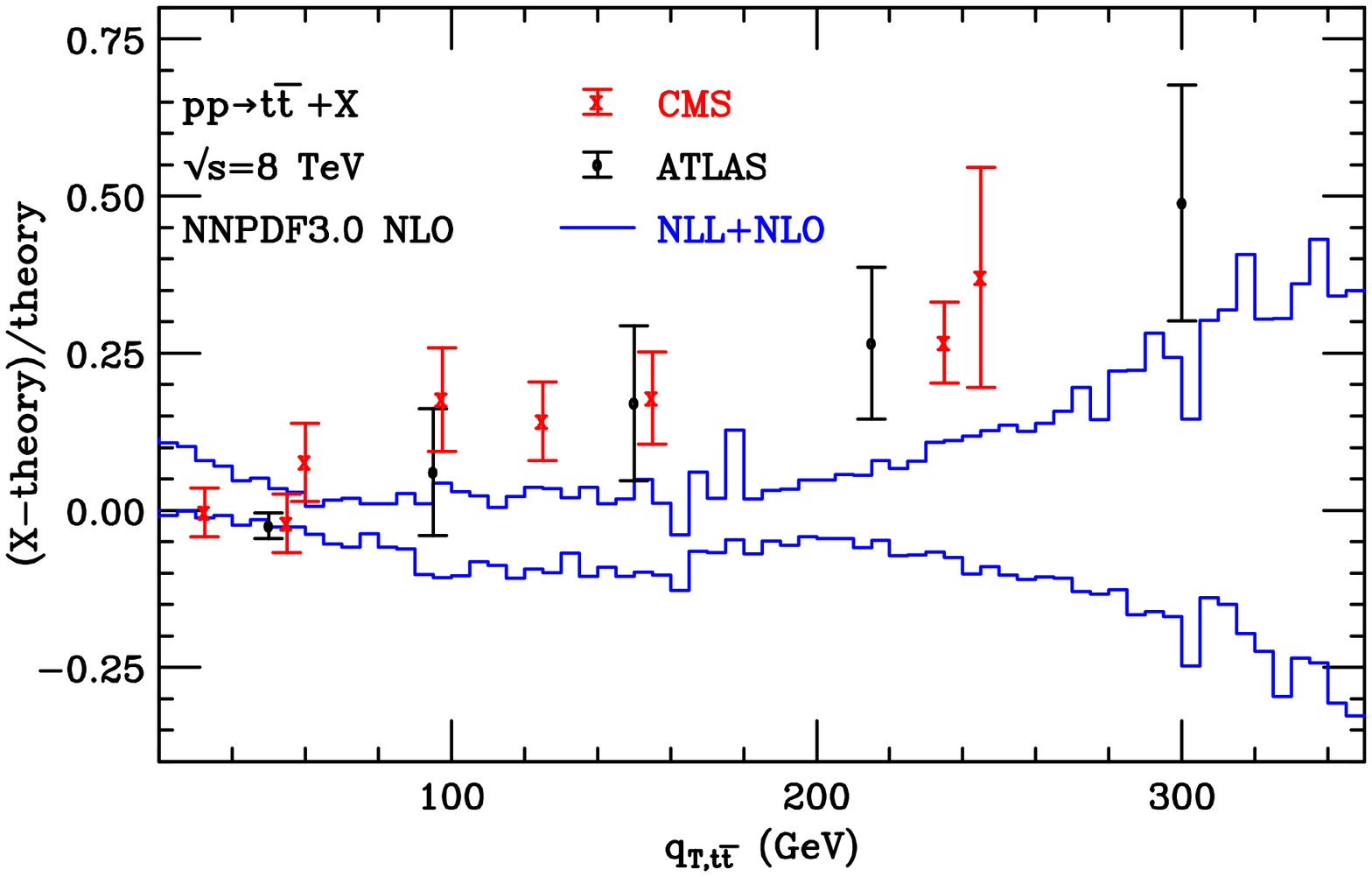}
}
\subfigure[]{
\includegraphics[width=0.3\textwidth,angle=90]{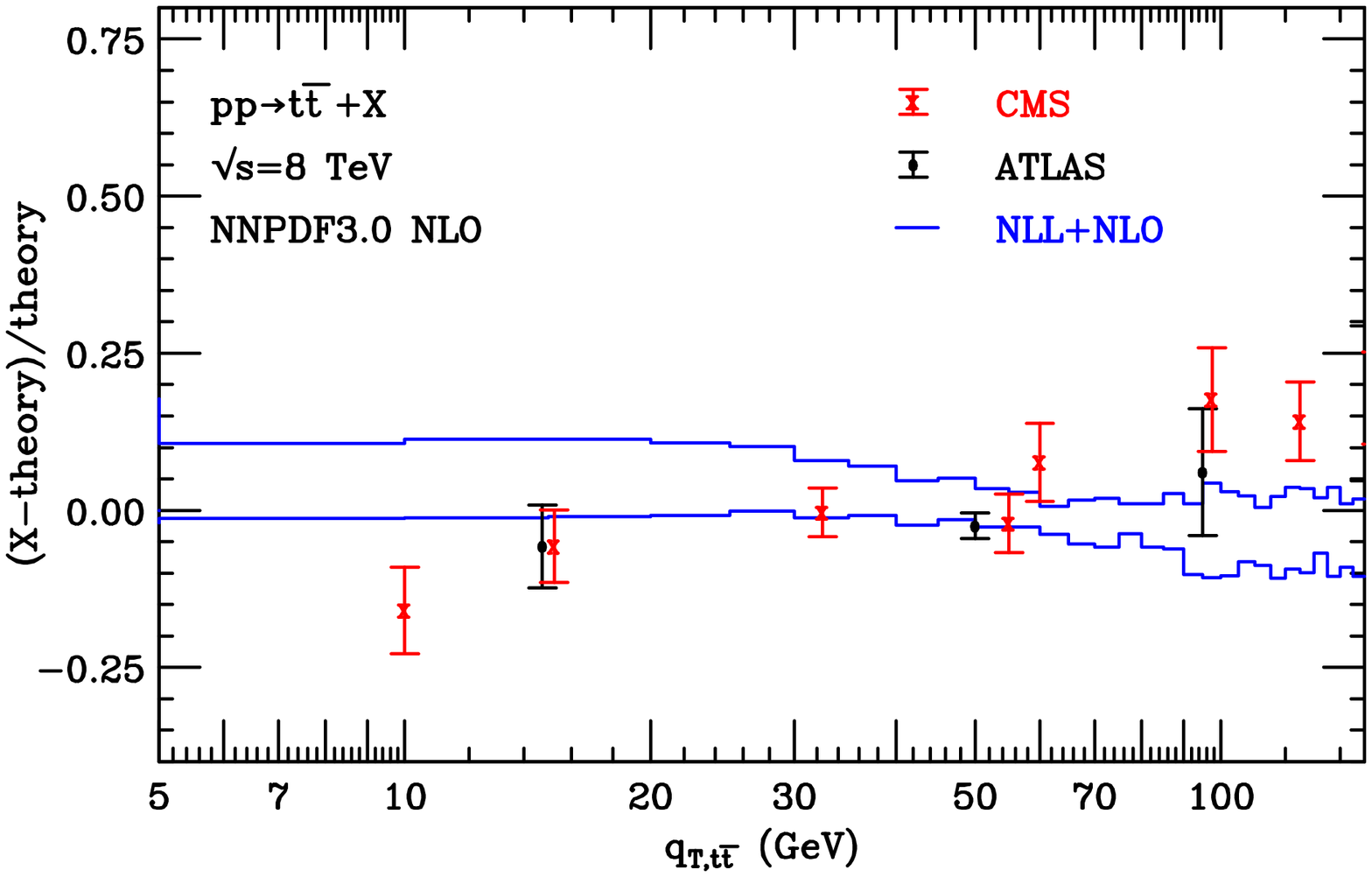}
}
\caption{\label{fig:NLL_data}
{\em Fractional difference of scale dependent NLL+NLO prediction
and LHC data \cite{Aaboud:2016iot,Khachatryan:2015oqa} 
with respect to the NLL+NLO result at $\mu_F=\mu_R=Q=m_t$.
The results refer to the normalized $q_T$ distribution $1/\sigma\,(d\sigma/dq_T)$
of $t{\bar t}$ pairs at ${\sqrt s}=8$~TeV.
The two panels highlight (a) the intermediate and large $q_T$ region and 
(b) the small-$q_T$ region.}}
\end{figure}

In Fig.~\ref{fig:NLL_data} we compare our resummed predictions at NLL+NLO accuracy
with the LHC data. The presentation of the results in Fig.~\ref{fig:NLL_data} is analogous to that of the corresponding fixed-order comparison in 
Fig.~\ref{fig:NLO_data}. The plots in Fig.~\ref{fig:NLL_data} show the 
fractional differences (X-`theory')/`theory', where X is either the NLL+NLO prediction with its scale uncertainty (which is computed as in the results of
Fig.~\ref{fig:X-NLO}) or the LHC data \cite{Aaboud:2016iot, Khachatryan:2015oqa},
and the reference theoretical result (`theory') is the NLL+NLO prediction at central scales ($\mu_F=\mu_R=Q=m_t$). As in the case of Fig.~\ref{fig:NLO_data}, the NLL+NLO
results are obtained by using $q_T$ bins
with a constant size of 5~GeV,
while the comparison to the data is done by using exactly the same bin sizes that are used in the experimental measurements.
The bin-by-bin comparison between data and NLL+NLO results is explicitly reported also
in Tables~\ref{tab:atlas}, \ref{tab:cmsjets} and \ref{tab:cmsdilep}.

The region of small and intermediate values of $q_T$ is highlighted in 
Fig.~\ref{fig:NLL_data}-right.
Throughout the low-$q_T$ region the NLL+NLO result has a scale uncertainty that is uniformly of the order of about $\pm 10\%$. 
This should be contrasted with the behaviour at fixed order (see Fig.~\ref{fig:NLO_data}-right), since the (NLO) NNLO scale uncertainty dramatically increases as $q_T\to 0$.
In the low-$q_T$ region we also see (Fig.~\ref{fig:NLL_data}-right and Tables~\ref{tab:atlas}--\ref{tab:cmsdilep}) that 
the NLL+NLO prediction is consistent with the data within the corresponding uncertainties.
Only the first bin of the CMS lepton+jets measurement (Table~\ref{tab:cmsjets})
is outside the scale uncertainty band. In the intermediate-$q_T$ region
(say, in the range 50~GeV$\ltap q_T \ltap 150$~GeV) data and NLL+NLO results are consistent. The degree of consistency is comparable to that between data and NNLO results (Fig.~\ref{fig:NLO_data}). This is a consequence of the fact 
(see comments on Fig.~\ref{fig:X-NLO})
that NLL+NLO and NNLO results behave similarly at intermediate values of $q_T$.

At high values of $q_T$ (Fig.~\ref{fig:NLL_data}-left) the data tend to 
systematically overshoot the NLL+NLO results. As we have previously noticed, in this $q_T$ region the resummed calculation cannot improve the predictivity of the fixed-order expansion, which has to be preferred to obtain QCD predictions. At high $q_T$, the agreement with the data definitely improves by replacing the NLL+NLO results with the fixed-order results at NNLO accuracy (Fig.~\ref{fig:NLO_data}-left).

\begin{table}
\centering
\begin{tabular}{| c | c |  c |}
\hline
  $q_{T,t\bar{t}}\,[\mathrm{GeV}]$ & $\frac{1}{\sigma}\frac{d\sigma}{dq_{T,t\bar{t}}}\,[\mathrm{TeV}^{-1}]$ ATLAS & $\frac{1}{\sigma}\frac{d\sigma}{dq_{T,t\bar{t}}}\,[\mathrm{TeV}^{-1}]$ NLL+NLO \\
  \hline
  0-30 & 14.3 $\pm$ 1.0 & $15.2 \substack{+1.4 \\ -0.3}$  \\
  \hline
  30-70 & 7.60 $\pm$ 0.16 & $7.79 \substack{+0.38 \\ -0.17}$ \\
  \hline
  70-120 & 2.94 $\pm$ 0.28 & $2.77 \substack{+0.05 \\ -0.21}$ \\
  \hline
  120-180 & 1.14 $\pm$ 0.12 & $0.97 \substack{+0.03 \\ -0.09}$ \\
  \hline
  180-250 & 0.42 $\pm$ 0.04 &  $0.33 \substack{+0.02 \\ -0.02}$ \\
  \hline
  250-350 & 0.143 $\pm$ 0.018 & $0.096 \substack{+0.021 \\ -0.015}$ \\
  \hline
\end{tabular}
\caption{\label{tab:atlas}
{\em The normalized $q_T$ distribution $1/\sigma\,(d\sigma/dq_T)$
of $t{\bar t}$ pairs in $pp$ collisions at the LHC (${\sqrt s}=8$~TeV):
comparison between ATLAS data (dilepton channels) \cite{Aaboud:2016iot} and NLL+NLO results.}} 
\end{table}

\begin{table}
\centering
\begin{tabular}{| c | c |  c |}
\hline
  $q_{T,t\bar{t}}\,[\mathrm{GeV}]$ & $\frac{1}{\sigma}\frac{d\sigma}{dq_{T,t\bar{t}}}\,[\mathrm{TeV}^{-1}]$ CMS & $\frac{1}{\sigma}\frac{d\sigma}{dq_{T,t\bar{t}}}\,[\mathrm{TeV}^{-1}]$ NLL+NLO \\
  \hline
  0-20 & 13.2 $\pm$ 1.1 & $15.7 \substack{+1.9 \\ -0.2}$ \\
  \hline
  20-45 & 11.8 $\pm$ 0.5 & $11.8 \substack{+1.0 \\ -0.1}$ \\
  \hline
  45-75 & 6.40 $\pm$ 0.37 & $5.95 \substack{+0.17 \\ -0.20}$ \\
  \hline
  75-120 & 2.84 $\pm$ 0.20 & $2.41 \substack{+0.23 \\ -0.02}$ \\
  \hline
  120-190 & 1.07 $\pm$ 0.07 & $0.91 \substack{+0.03 \\ -0.08}$ \\
  \hline
  190-300 & 0.306 $\pm$ 0.039 & $0.223 \substack{+0.022 \\ -0.018}$ \\
  \hline
\end{tabular}
\caption{\label{tab:cmsjets}
{\em The normalized $q_T$ distribution $1/\sigma\,(d\sigma/dq_T)$
of $t{\bar t}$ pairs in $pp$ collisions at the LHC (${\sqrt s}=8$~TeV):
comparison between CMS data (lepton+jets channels) \cite{Khachatryan:2015oqa} 
and NLL+NLO results.}} 
\end{table}

\begin{table}
\centering
\begin{tabular}{| c | c |  c |}
\hline
  $q_{T,t\bar{t}}\,[\mathrm{GeV}]$ & $\frac{1}{\sigma}\frac{d\sigma}{dq_{T,t\bar{t}}}\,[\mathrm{TeV}^{-1}]$ CMS & $\frac{1}{\sigma}\frac{d\sigma}{dq_{T,t\bar{t}}}\,[\mathrm{TeV}^{-1}]$ NLL+NLO \\
  \hline
  0-30 & 14.3 $\pm$ 0.9 & $15.2 \substack{+1.8 \\ -0.2}$ \\
  \hline
  30-80 & 6.9 $\pm$ 0.3 & $7.0 \substack{+0.3 \\ -0.2}$ \\
  \hline
  80-170 & 1.91 $\pm$ 0.11 & $1.67 \substack{+0.04 \\ -0.15}$ \\
  \hline
  170-300 & 0.347 $\pm$ 0.018 & $0.274 \substack{+0.023 \\ -0.002}$ \\
  \hline
\end{tabular}
\caption{\label{tab:cmsdilep}
{\em The normalized $q_T$ distribution $1/\sigma\,(d\sigma/dq_T)$
of $t{\bar t}$ pairs in $pp$ collisions at the LHC (${\sqrt s}=8$~TeV):
comparison between CMS data (dilepton channels) \cite{Khachatryan:2015oqa} 
and NLL+NLO results.}} 
\end{table}

In Sect.~\ref{sec:theo} we have recalled and noticed that a 
distinctive feature of
transverse-momentum resummation for heavy-quark production is the appearance 
of dynamical colour-correlation effects
in the resummed form factor $\exp{\cal G}$ (see Eq.~(\ref{eq:ff})).
These effects are due to soft-parton radiation from the $t{\bar t}$ pair, and they start to contribute at the NLL level through a soft wide-angle component of the function $g_N^{(2)}$ in Eq.~(\ref{eq:gexp}). It is of interest to quantify the impact
of these effects. To this purpose we consider the NLL+NLO calculation of the spectrum by removing the contribution of the soft wide-angle component to the resummed 
exponent of Eq.~(\ref{eq:gexp}) (this is equivalent to set ${\bf \Gamma}_t^{(1)}=0$,
${\bf \Gamma}_t^{(1)}$ being the one-loop soft anomalous dimension of 
Ref.~\cite{Catani:2014qha}). We note, however, that the regular component 
$d\sigma^{\rm (reg)}$ (see Eq.~(\ref{eq:singreg})) of the NLL+NLO calculation
is left unchanged, so as not to spoil the matching procedure with the complete NLO
result at low values of $q_T$ (removing the effect of ${\bf \Gamma}_t^{(1)}$ from
$d\sigma^{\rm (reg)}/dq_T$ would lead to a divergent cross section in the limit $q_T \to 0$).

\begin{figure}[th]
\centering
\includegraphics[width=0.58\textwidth,angle=90]{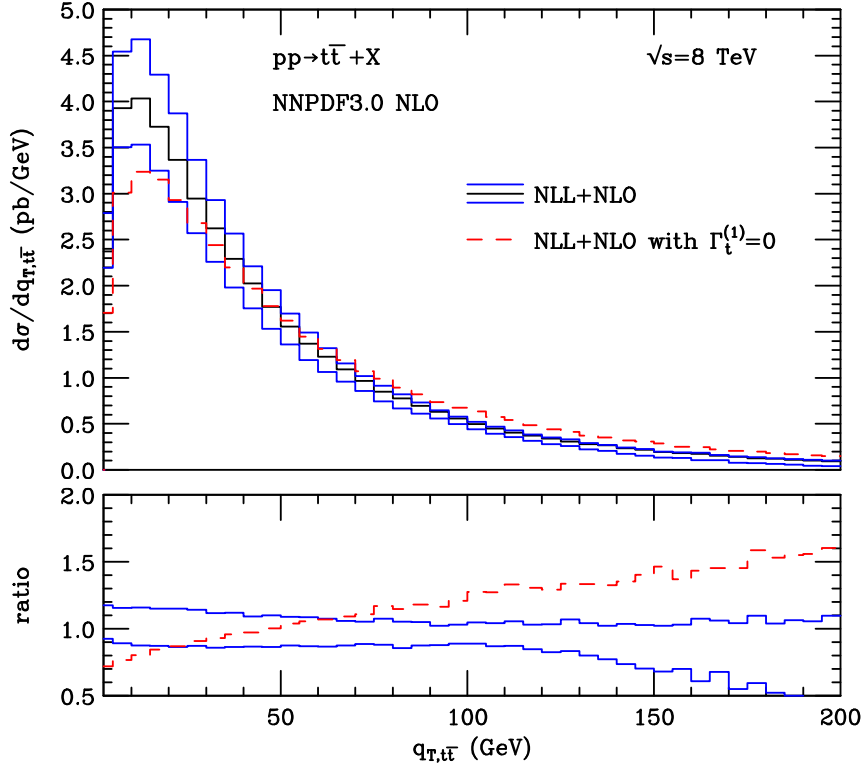}\hfill
\caption{\label{fig:soft}
{\em The transverse-momentum cross section $d\sigma/dq_T$ of the $t\bar{t}$ pair at the LHC ($\sqrt{s}=8~\mathrm{TeV}$). The scale dependent result at NLL+NLO accuracy
(solid) is compared with the `modified' result (dashed) that is obtained by removing the resummed contribution of soft wide-angle radiation (which is given by the soft anomalous dimension ${\bf \Gamma}_t^{(1)}$) at central scales ($\mu_R=\mu_F=Q=m_t$).
The lower panel presents the ratios of the NLL+NLO scale dependence and  of the `modified' result with respect to the NLL+NLO result at central scales.}}
\end{figure}

The modified (by setting ${\bf \Gamma}_t^{(1)}=0$ in $g_N^{(2)}$) NLL+NLO result
for $d\sigma/dq_T$ at central values of the scales ($\mu_F=\mu_R=Q=m_t$) is presented in Fig.~\ref{fig:soft} (dashed lines), where it is compared to the full NLL+NLO result with its scale dependence (solid lines). Since our modification only 
affects the function $g_N^{(2)}$ in the resummed form factor $\cal G$, the integral
over $q_T$ of the dashed histogram in Fig.~\ref{fig:soft} coincides with the NLO total cross section. Therefore, at central values of the scales, the solid and dashed histograms in Fig.~\ref{fig:soft} differ only in their shape, and the $q_T$ shape
of solid histogram is softer than that of the dashed histogram. 
As we can see, the effect of the soft wide-angle radiation
(which is included in the full NLL+NLO result) is positive at small values of $q_T$:
in the lowest-$q_T$ bin, its size is at the level of about $+ 40\,\%$, which is well outside the range of the scale variation band. At larger values of $q_T$
($q_T \gtap 50$~GeV), the effect is negative and it reaches the size of about 
$-40\,\%$ at $q_T \sim 200$~GeV.

We can present a qualitative interpretation of the results in Fig.~\ref{fig:soft}.
The soft anomalous dimension ${\bf \Gamma}_t^{(1)}$ formally parametrizes the differences at the resummed level between heavy-quark production and the production of a colourless system from the same colliding partons. The term 
${\bf \Gamma}_t^{(1)}$ is due to soft radiation at large angles with respect to the direction of the initial-state colliding partons (hadrons). Soft wide-angle radiation receives two physically distinct contributions: final-state radiation from the heavy-quark pair and initial/final-state interferences due to the non-vanishing colour charge of the heavy quarks. Final-state radiation tends to make the $q_T$ spectrum harder. Owing to colour-coherence effects, interferences have instead a destructive 
character and they soften the $q_T$ spectrum (a related discussion of colour-coherence in the context of threshold resummation is presented in Sect.~4 of 
Ref.~\cite{Catani:2013vaa}).
In the case of $t{\bar t}$ production the colour-coherence destructive interferences
dominate with respect to the final-state radiation effect, since the latter is suppressed (screened) by the large value of the top-quark mass. It follows that the presence of ${\bf \Gamma}_t^{(1)}$ leads to a softer $q_T$ spectrum in qualitative agreement with the results in Fig.~\ref{fig:soft}.
This is the overall effect of colour-coherence on the $q_T$ spectrum. 
As recalled in Sect.~\ref{sec:theo}, the colour correlation effects due to 
${\bf \Gamma}_t^{(1)}$ have a definite dynamical dependence on the quark (or antiquark) scattering angle $\theta$. Detailed studies of the $q_T$ cross section at fixed values of the scattering angle can be useful to futher investigate colour-coherence effects in $t{\bar t}$ production at small values of $q_T$.

\begin{figure}[th]
\centering
\includegraphics[width=0.58\textwidth,angle=90]{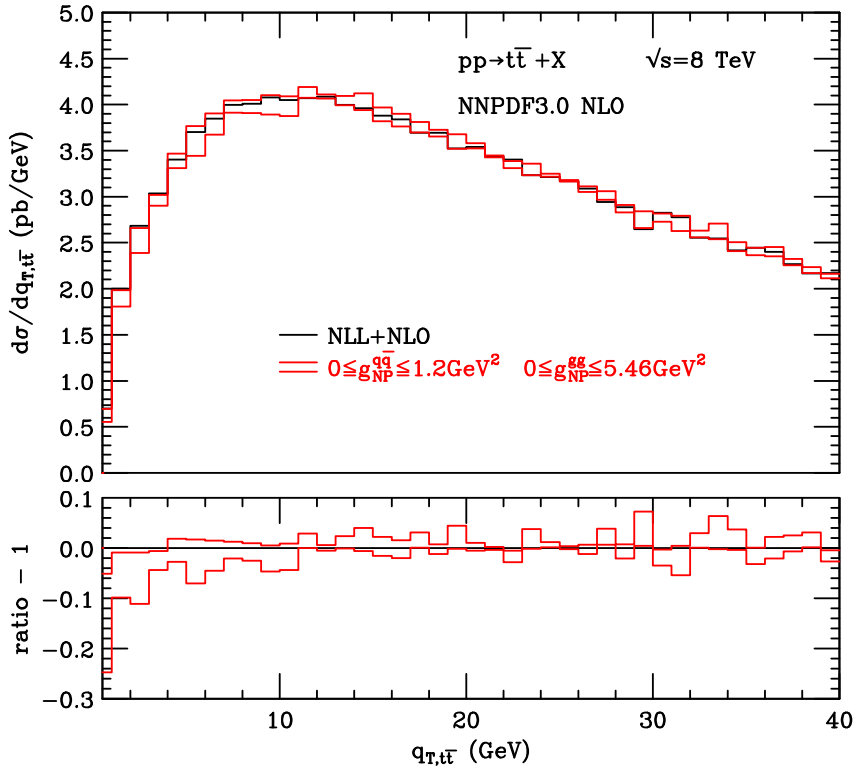}\hfill
\caption{\label{fig:NP}
{\em The transverse-momentum cross section $d\sigma/dq_T$ of the $t\bar{t}$ pair at the LHC ($\sqrt{s}=8~\mathrm{TeV}$). Comparison of NLL+NLO results
at central scales ($\mu_R=\mu_F=Q=m_t$) without (black line) and with (red band)
the inclusion of a NP form factor 
$(0\leq g_{\mathrm{NP}}^{q\bar{q}}\leq 1.2\,\mathrm{GeV}^2$, $0\leq g_{\mathrm{NP}}^{gg}\leq 5.46\,\mathrm{GeV}^2)$.
The lower panel presents the relative effect of the NP form factor.}}
\end{figure}

The resummed predictions that we have presented so far are obtained
in a purely perturbative
framework. As briefly recalled at the end of Sect.~\ref{sec:theo},
the transverse-momentum distribution of high-mass
systems is affected by NP effects, which may become significant as $q_T\to 0$.
To check the impact of NP effects at small values of $q_T$, we use the customary procedure of multiplying the resummation component of the cross section in $b$ space
(see Eq.~(\ref{eq:ff}))
by a NP smearing form factor $S_{NP}$ of gaussian form:
\begin{equation}
S_{NP}=\mathrm{exp}\left\{-g_{NP}
\,b^2\right\}\,.
\end{equation}
We use different values of the NP parameter $g_{NP}$ for the $q\bar{q}$ annihilation and $gg$ fusion channels (see Eq.~(\ref{eq:resw})).
Specifically, we vary $g_{NP}$ in the range $0\leq g_{\mathrm{NP}}^{q\bar{q}}\leq 1.2\,\mathrm{GeV}^2$ for the $q\bar{q}$ driven contribution, and in the range $0\leq g_{\mathrm{NP}}^{gg}\leq 5.46\,\mathrm{GeV}^2$ for the $gg$ contribution. 
These ranges of $g_{NP}$ are those that were considered
in Ref.~\cite{Catani:2015vma} and Ref.~\cite{deFlorian:2011xf} for
vector and Higgs boson production, respectively.
The quantitative effects of the NP form factor on the NLL+NLO result for 
$d\sigma/dq_T$ are shown in Fig.~\ref{fig:NP}.  
The red band corresponds to the envelope of the NP results that are obtained 
by varying $g^{q\bar{q}}_{NP}$ and $g^{gg}_{NP}$ in their specified ranges.
We see that the NP effects are small and, in particular, definitely smaller than
the perturbative uncertanties due to scale variations. This conclusion holds with the exception of the region of very small values of $q_T$ (say, $q_T\ltap 5$ GeV), where the impact of NP effects can of the order of $10\%$. This result is consistent with analogous findings in the case
of vector \cite{Catani:2015vma} and Higgs \cite{deFlorian:2011xf} boson production.

As stated at the beginning of Sect.~\ref{sec:res}, our resummed results are obtained by using the value $b_{\rm lim}=3~{\rm GeV}^{-1}$ of the parameter $b_{\rm lim}$
in Eq.~(\ref{bstar}). We have examined variations of $b_{\rm lim}$ by about a factor of two and we find quantitative effects that are definitely smaller than those produced by the NP form factor.

In this paper we do not study how different PDF sets affect the QCD predictions for the $q_T$ spectrum of the $t{\bar t}$ pair. A study of the normalized $q_T$ distribution $1/\sigma (d\sigma/dq_T)$ at ${\sqrt s}=8$~TeV by using different PDF sets is performed by the ATLAS Collaboration in Ref.~\cite{Aaboud:2016iot}. The results in Ref.~\cite{Aaboud:2016iot}
(see, e.g., Fig.~6b therein) show that various sets of modern PDFs produce differences
that are below the 10\% level.

Our NLL+NLO calculation consistently combines the NLO perturbative result, which is well-behaved at high values of $q_T$, with the resummation of the 
logarithmically-enhanced terms in the small-$q_T$ region. 
Using the resummation formalism in impact parameter space
(see Eq.~(\ref{eq:ff})) we achieve control of the large logarithmic terms in analytic form up to NLL accuracy.
Parton shower (PS) Monte Carlo event generators provide an alternative way for carrying out the resummation procedure.
In this case the resummation of the large logarithmic terms is effectively achieved through the PS,
although within the limited logarithmic accuracy
of present PS algorithms.
The matching of the PS with the exact NLO calculation, which leads to 
NLO+PS generators,
can be carried out by using
the {\sc MC@NLO} \cite{Frixione:2002ik,Frixione:2003ei} or {\sc POWHEG} \cite{Nason:2004rx,Frixione:2007vw} methods.
Both methods have been applied to $t{\bar t}$ hadroproduction
\cite{Frixione:2003ei,Frixione:2007nw}.

It is of interest to present an illustrative comparison between our NLL+NLO results
and those of a NLO+PS Monte Carlo generator. We choose a Monte Carlo generator based on the
{\sc POWHEG} method, and we use the {\sc POWHEG} {\sc BOX} implementation 
\cite{Alioli:2010xd} to interface the $t{\bar t}$ NLO calculation to the 
version 6.4.25 of the {\sc PYTHIA} parton shower
\cite{Sjostrand:2006za} with the default values of the parameters.
The {\sc POWHEG}+{\sc PYTHIA} calculation is performed by varying the parameter
$h_{\rm damp}$ \cite{Alioli:2008tz}. Within the {\sc POWHEG} method, the radiation
of the final-state parton with the largest transverse-momentum $p_T$
(`the hardest radiation') is generated at first (i.e., before parton showering effects) according to the NLO result and it is weighted by a Sudakov form factor that is obtained by exponentiating a fraction $F$ of the NLO real-emission contribution.
The fraction $F$ is controlled by the parameter $h_{\rm damp}$ in the form
\begin{equation}
F(p_T)=\f{h^2_{\rm damp}}{h^2_{\rm damp}+p_T^2}\, ,
\end{equation}
so that larger values of $h_{\rm damp}$ are expected to produce a harder $q_T$ spectrum of the $t{\bar t}$ pair.

\begin{figure}[th]
\centering
\includegraphics[width=0.58\textwidth,angle=90]{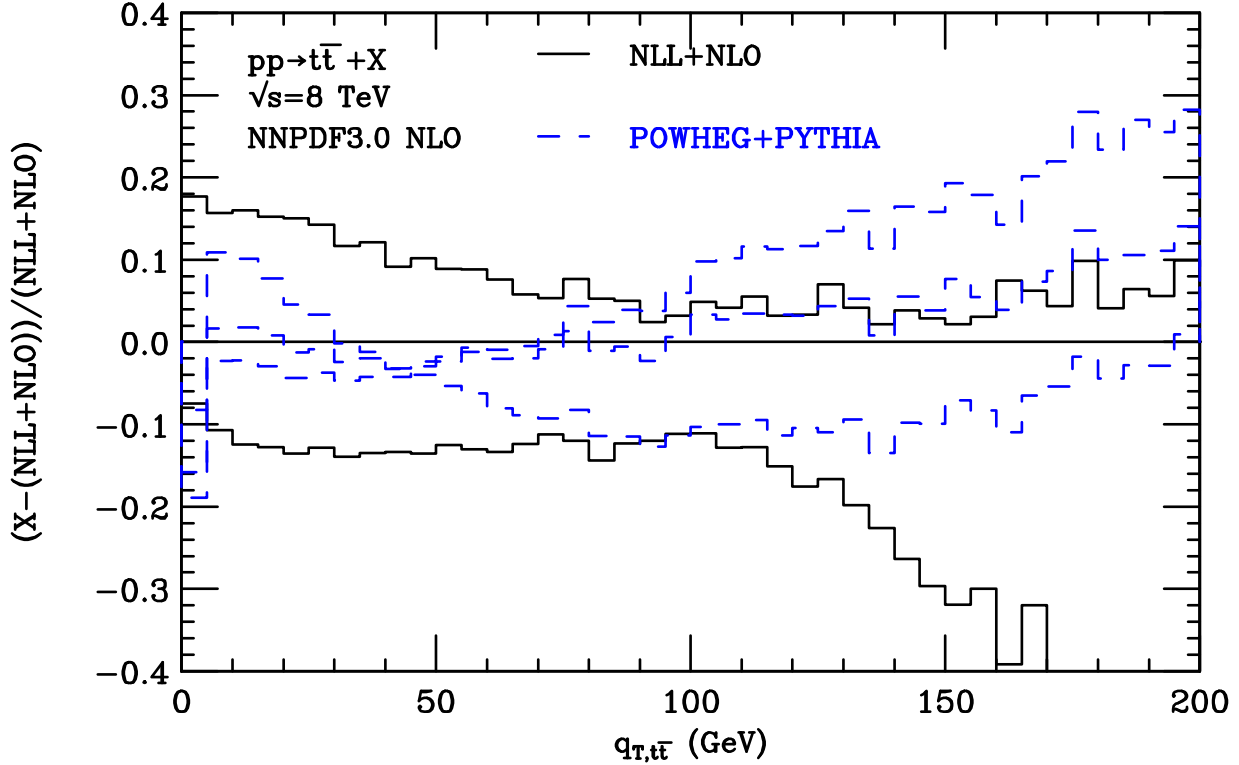}\hfill
\caption{\label{fig:nll_powheg}
{\em Fractional difference of the POWHEG+PYTHIA results (dashed) with respect to the NLL+NLO result at central scales ($\mu_F=\mu_R=Q=m_t$). The results refer to the differential cross section $d\sigma/dq_T$ of $t{\bar t}$ pairs at ${\sqrt s}=8$~TeV.
The POWHEG+PYTHIA results are presented for three different values of $h_{\rm damp}$,
$h_{\rm damp}=\{ m_t/2, m_t, 2m_t \}$, at fixed $\mu_F=\mu_R=m_t$. The scale dependence of the NLL+NLO results
(solid) is also reported.
}}
\end{figure}

The comparison between our resummed results and those obtained by using 
{\sc POWHEG} + {\sc PYTHIA} is presented in Fig.~\ref{fig:nll_powheg}. We consider the differential cross section $d\sigma/dq_t$ for $t{\bar t}$ production at the LHC
(${\sqrt s}=8$~TeV), and in this Figure we present the corresponding fractional difference (X-`theory')/`theory'. The reference theoretical result (`theory') is
the NLL+NLO calculation at central values of the scales ($\mu_R=\mu_F=Q=m_t$), and X denotes either the NLL+NLO scale dependence 
(computed as in Figs.~\ref{fig:X-NLO} and \ref{fig:soft}) or the
{\sc POWHEG}+{\sc PYTHIA} result at three different values of $h_{\rm damp}$ 
($h_{\rm damp}=\{m_t/2,m_t,2m_t\}$) and fixed $\mu_R=\mu_F=m_t$.
The comparison in Fig.~\ref{fig:nll_powheg} shows that the NLL+NLO result and the 
{\sc POWHEG}+{\sc PYTHIA} result are consistent with each other within their uncertainties, and they are quite similar at small and intermediate values of $q_T$.

We note that the $h_{\rm damp}$ dependence computed in Fig.~\ref{fig:nll_powheg}
certainly corresponds to a lower limit on the uncertainty of 
{\sc POWHEG}+{\sc PYTHIA}. Indeed, there are various other sources of theoretical uncertainties in NLO+PS event generators (see, e.g., Ref.~\cite{Nason:2012pr}).
These include effects from variations of $\mu_R$ and $\mu_F$, from the detailed matching procedure with the NLO calculation and from the uncertainties of the PS algorithms.

In Refs.~\cite{Aaboud:2016iot} and \cite{Khachatryan:2015oqa}
ATLAS and CMS data on the $q_T$ spectrum of $t{\bar t}$ pairs at the LHC
(${\sqrt s}=8$~TeV) have been compared with the results of various NLO+PS event generators, which use both the {\sc MC@NLO} and {\sc POWHEG} methods interfaced to
{\sc PYTHIA} and {\sc HERWIG} \cite{Corcella:2000bw} for parton showering.
The comparison shows that the measured $q_T$ spectrum is well described 
\cite{Aaboud:2016iot,Khachatryan:2015oqa}
by these Monte Carlo event generators. Owing to the agreement between these data and our NLL+NLO results (as previously discussed), it turns out that the NLL+NLO results and those of NLO+PS generators are relatively similar at small and intermediate values of $q_T$.

This similarity is partly expected. The PS algorithm effectively resums LL contributions in the small-$q_T$ region. Some contributions at the NLL level are also included through the matching with the NLO result and through the PS. In particular,
the PS produces final-state radiation from the $t{\bar t}$ pair and it can take into account soft-gluon colour-coherence effects (though they are usually approximated in the large-$N_c$ limit, $N_c$ being the number of colours) \cite{Marchesini:1989yk}
by using angular-ordered evolution or dipole showering.

As recalled in Sect.~\ref{sec:intro}, quantitative results of a resummed calculation for the $q_T$ spectrum of $t{\bar t}$ pairs at LHC energies were obtained in 
Refs.~\cite{Zhu:2012ts, Li:2013mia}. The resummed calculation of 
Ref.~\cite{Li:2013mia} and our calculation at NLL+NLO accuracy are based on equivalent formal inputs \cite{Li:2013mia, Catani:2014qha}, but they use different implementation formalisms. The differences regard, for instance, the type of auxiliary scales that are used to organize the resummed calculation, a partly different classification of the resummed logarithmic terms and the inclusion in 
Ref.~\cite{Li:2013mia} of some beyond-NLL effects due to the soft anomalous dimension
${\bf \Gamma}_t^{(2)}$ at ${\cal O}(\as^2)$ \cite{Li:2013mia, Catani:2014qha}.
In view of these implementation differences, ensuing quantitative differences between our results and those of Ref.~\cite{Li:2013mia} are expected. The results of
Ref.~\cite{Li:2013mia} were compared to ATLAS and CMS data in 
Refs.~\cite{Aaboud:2016iot, Khachatryan:2015oqa}. From that comparison and, in particular, from the Theory/Data plots in Fig.~9 of Ref.~\cite{Aaboud:2016iot}
and Figs.~11 and 14 of Ref.~\cite{Khachatryan:2015oqa} we can infer an indirect comparison with our results. We find that the $q_T$ spectrum of Ref.~\cite{Li:2013mia}
tends to be softer than our spectrum, and still softer than the spectrum  of the ATLAS and CMS data. More precisely, in the region of small and medium values of 
$q_T$ (where, say, $q_T \ltap 120$~GeV), the central prediction of 
Ref.~\cite{Li:2013mia} is consistent with our results within our scale uncertainty band. At larger values of $q_T$, the central prediction of Ref.~\cite{Li:2013mia}
lies outside our scale uncertainty band, and it sizeably deviates from both the NLO result and the data (as pointed out in Refs.~\cite{Aaboud:2016iot, Khachatryan:2015oqa}).

\section{Summary}
\label{sec:sum}

In this paper we have considered the transverse-momentum cross section of $t{\bar t}$ pairs at the LHC.

We have first presented fixed-order predictions at NLO (i.e., up to ${\cal O}(\as^3)$) and NNLO (i.e., up to ${\cal O}(\as^4)$) and we have compared them with ATLAS and CMS data at $\sqrt{s}=8$~TeV.
At intermediate and large values of $q_T$ the NNLO corrections improve the agreement with the data and the NNLO uncertainties are comparable to the experimental uncertainties.

At small values of $q_T$ the reliability of the fixed-order perturbative expansion
is spoiled by the presence of large logarithmic contributions. These logarithmic contributions have to be resummed to all perturbative orders.
Using the resummation formalism of Ref.~\cite{Catani:2014qha} we have presented quantitative results up to NLL+NLO accuracy.
The NLL+NLO result has uniform theoretical accuracy throughout the regions of small and intermediate values of $q_T$, leading to a perturbative uncertainty of the order of 
$\pm 10\%$ for the normalised $q_T$ distribution.
In these regions 
the resummed result is consistent with the ATLAS and CMS data at $\sqrt{s}=8$~TeV within the corresponding uncertainties.
At large values of $q_T$ the resummation procedure cannot improve the predictivity of the fixed-order expansion.
The NNLO resummation contribution ${\cal H}^{(2)}$ (see Eq.~(\ref{eq:hfun})) is not yet known. Once ${\cal H}^{(2)}$ becomes available, theoretical improvements of the resummed predictions can be obtained by extending the resummed calculation to 
NNLL+NNLO accuracy.

The NLL+NLO result for the $q_T$ spectrum receives resummed contributions
from soft wide-angle radiation through a corresponding soft anomalous dimension.
We have examined the quantitative impact of these contributions and we have shown that they
make the $q_T$ spectrum softer, in qualitative agreement with 
a discussion of colour-coherence effects.
We have finally presented a comparison of our resummed calculation with the prediction of the Monte Carlo event generator {\sc POWHEG}+{\sc PYTHIA}. We find that the two predictions are consistent within their estimated uncertainties.

\paragraph{Acknowledgements.} We are grateful to Stefan Kallweit for his help with the {\sc Munich} code. This research was supported in part by the Swiss National Science Foundation (SNF) under contract 200020-169041 and by the Research Executive Agency (REA) of the 
European Union under the Grant Agreement number PITN-GA-2012-316704 ({\it Higgstools}).

\renewcommand{\refname}{

\end{document}
\begin{thebibliography}{99}
\bibitem{Kallweit:Munich}
S.~Kallweit,
{\sc Munich} is the abbreviation of ``MUlti-chaNnel Integrator at Swiss~(CH) precision''---an automated parton level NLO 
generator. 
In preparation.
\end{thebibliography}
